\documentclass[useAMS,usenatbib]{mn2e}

\usepackage{times}
\usepackage{epsfig, amssymb}
\usepackage{graphics}
\usepackage{graphicx}
\usepackage{eufrak}
\usepackage{mathrsfs}
\usepackage{eucal}
\usepackage{amsmath}
\usepackage{fontenc}

\def\mnras{{MNRAS}}

\def\etal{et al.~}  

\newcommand{\ergsec}{\ifmmode \mathrm{erg~s}^{-1} \else erg~s$^{-1}$\fi}
\newcommand{\Msun}{\ifmmode ~\mathrm M_{\odot} \else $~\mathrm M_{\odot}$\fi}
\newcommand{\Mbh}{\ifmmode M_{\mathrm{BH}} \else $M_{\mathrm{BH}}$\fi}
\newcommand{\Mbulge}{\ifmmode M_{\mathrm{Bulge}} \else $M_{\mathrm{Bulge}}$\fi}
\newcommand{\Medd}{\ifmmode \dot{M}_{\mathrm{Edd}} \else $\dot{M}_{\mathrm{Edd}}$\fi}
\newcommand{\Lbol}{\ifmmode L_{\mathrm{bol}} \else $L_{\mathrm{bol}}$\fi}
\newcommand{\Ledd}{\ifmmode L_{\mathrm{Edd}} \else $L_{\mathrm{Edd}}$\fi}
\newcommand{\Lsun}{\ifmmode L_{\odot} \else $L_{\odot}$\fi}
\newcommand{\Lsx}{\ifmmode L_{\mathrm{SX}} \else $L_{\mathrm{SX}}$\fi}
\newcommand{\Mbj}{\ifmmode M_{b_{\rm J}} \else $M_{b_{\rm J}}$\fi}
\newcommand{\bj}{\ifmmode b_{\rm J} \else $b_{\rm J}$\fi}
\newcommand{\Lhx}{\ifmmode L_{\mathrm{HX}} \else $L_{\mathrm{HX}}$\fi}
\newcommand{\ledd}{\ifmmode \lambda_{\mathrm{Edd}} \else $\lambda_{\mathrm{Edd}}$\fi}
\newcommand{\lgledd}{\ifmmode \log\lambda_{\mathrm{Edd}} \else $\log\lambda_{\mathrm{Edd}}$\fi}
\newcommand{\rbh}{\ifmmode \rho_{\mathrm{BH}} \else $\rho_{\mathrm{BH}}$\fi}
\newcommand{\m}{\ifmmode \dot{m} \else $\dot{m}$\fi}
\newcommand{\fhalo}{\ifmmode f_{\mathrm{Halo}}^{\mathrm{act}} \else $f_{\mathrm{Halo}}^{\mathrm{act}}$\fi}
\newcommand{\fvis}{\ifmmode f_{\mathrm{vis}} \else $f_{\mathrm{vis}}$\fi}
\newcommand{\fobsc}{\ifmmode f_{\mathrm{obsc}} \else $f_{\mathrm{obsc}}$\fi}
\newcommand{\fq}{\ifmmode f_{\mathrm{q}} \else $f_{\mathrm{q}}$\fi}

\title[The evolution of AGN across cosmic time]{The evolution of AGN across cosmic time: what is downsizing?}
\author[Fanidakis \etal]{\newauthor N.  Fanidakis,$^{1}$ C. M. Baugh,$^{1}$ A. J. Benson,$^{2}$ R. G. Bower,$^{1}$ S. Cole,$^{1}$ C. Done,$^{1}$
\newauthor  C. S. Frenk$^{1}$, R. C. Hickox$^{1}$, C. Lacey$^{1}$, C. del P. Lagos$^{1}$\\
$^{1}$Institute for Computational Cosmology, Department of Physics, University of Durham,\\
Science Laboratories, South Road, Durham DH1 3LE, United Kingdom\\
$^{2}$Mail Code 130-33, California Institute of Technology, Pasadena, CA 91125, USA}
\begin{document}    

\maketitle
\label{firstpage}

\begin{abstract}

We use a coupled model of the formation and evolution of galaxies and black holes (BH) to study the evolution of active galactic nuclei (AGN) in a cold dark matter universe.  The model predicts the BH mass, spin and mass accretion history. BH mass grows via accretion triggered by discs becoming dynamically unstable or galaxy mergers (called the starburst mode) and accretion from quasi-hydrostatic hot gas haloes (called the hot-halo mode). By taking into account AGN obscuration, we obtain a very good fit to the observed luminosity functions (LF) of AGN (optical, soft and hard X-ray, and bolometric) for a wide range of redshifts ($0<z<6$). The model predicts a hierarchical build up of BH mass, with the typical mass of actively growing BHs increasing with decreasing redshift. Remarkably, despite this, we find downsizing in the AGN population, in terms of the differential growth with redshift of the space density of faint and bright AGN. This arises naturally from the interplay between the starburst and hot-halo accretion modes. The faint end of the LF is dominated by massive BHs experiencing quiescent accretion via a thick disc, primarily during the hot-halo mode. The bright end of the LF, on the other hand, is dominated by AGN which host BHs accreting close to or in excess of the Eddington limit during the starburst mode. The model predicts that the comoving space density of AGN peaks at $z\simeq3$, similar to the star formation history. However, when taking into account obscuration, the space density of faint AGN peaks at lower redshift ($z\lesssim2$) than that of bright AGN ($z\simeq2-3$). This implies that the cosmic evolution of AGN is shaped in part by obscuration.
\end{abstract}


\section{Introduction}

Reproducing the evolution of active galactic nuclei (AGN) is a crucial test of galaxy formation models. Not only are AGN an important component of the Universe, they are also thought to play a key role in the evolution of their host galaxies and their environment. For example, feedback processes which accompany the growth of black holes (BHs) during AGN activity are thought to influence star formation (SF) through the suppression of cooling flows in galaxy groups and clusters (\citealt{dalla_vecchia_2004,springel_2005a, croton_2006, bower_2006, hopkins_2006, thacker_2006, somerville_2008, lagos_2008}, see also \citealt{marulli_2008}). Similarly, energy feedback from AGN is a key ingredient in determining the X-ray properties of the intracluster medium \citep{bower_2008, mccarthy_2010}. Understanding these processes requires that the underlying galaxy formation model is able to track and reproduce the evolution of AGN. 

Early surveys demonstrated that quasi-stellar objects (QSOs) undergo significant evolution from $z\sim0$ up to $z\sim2-2.5$ \citep{schmidt_1983, boyle_1988, hewett_1993, boyle_2000}. Beyond $z \sim 2$, the space density of QSOs starts to decline \citep{warren_1994, schmidt_1995, fan_2001, wolf_2003}. Recently there has been much progress in pinning down the evolution of faint AGN in the optical. Using the 2-degree-field (2dF) QSO Redshift survey, the luminosity function (LF) was probed to around a magnitude fainter than the break to $z \sim 2$ \citep[2QZ][]{croom_2001,croom_2004}. This limit was extended by the 2dF-SDSS luminous red galaxy and QSO (2SLAQ) survey \citep{richards_2005, croom_2009a}. The estimate of the LF from the final 2SLAQ catalogue reached $\Mbj\simeq-19.8$ at $z=0.4$ \citep{croom_2009b}. With the 2SLAQ LF, \citet{croom_2009b} demonstrated that in the optical, faint quasars undergo mild evolution, with their number density peaking at lower redshifts than is the case for bright quasars \citep[see also][]{bongiorno_2007}. 

Faint AGN can be selected robustly in X-rays \citep[][see also the review by \citealt{brandt_2005}]{hasinger_2001, giacconi_2002, alexander_2003, barcons_2007}, which means that a wider range of AGN luminosity can be probed in X-rays than is possible in the optical. This permits the study of the evolution of a wider variety of AGN in addition to QSOs (e.g., Seyfert galaxies) and thus provides a more representative picture of the various AGN populations. The evolution of the AGN LF in X-rays has been investigated by many authors by employing data from the Chandra, ASCA, ROSAT, HEAO-1 and XMM-Newton surveys \citep{miyaji_2000, la_franca_2002, cowie_2003, fiore_2003, ueda_2003, barger_2005, hasinger_2005}. These studies show that in both soft ($0.5-2\rm{keV}$) and hard ($2-10\rm{keV}$) X-rays, faint AGN are found to evolve modestly with redshift. In contrast, bright AGN show strong evolution, similar to that seen for quasars in the optical. In addition, observations in soft X-rays suggest that the comoving space density of bright AGN peaks at higher redshifts ($z\sim2$) than faint AGN \citep[$z<1$,][]{hasinger_2005}. 

The differential evolution of bright and faint AGN with redshift has been described as \emph{downsizing} \citep{barger_2005, hasinger_2005}. This implies that AGN activity in the low-$z$ universe is dominated by either high-mass BHs accreting at low rates or low-mass BHs growing rapidly. \citet{hopkins_2005b} proposed that the faint end of the LF is composed of high mass BHs experiencing quiescent accretion \citep[see also][]{hopkins_2005a, hopkins_2005c, babic_2007}. The bright end, on the other hand, in this picture corresponds to BHs accreting near their Eddington limit. In the \citeauthor{hopkins_2005b} model, quasar activity is short-lived and is assumed to be driven by galaxy mergers \citep{dimatteo_2005}. 

The mass of accreting BHs can be estimated using the spectra of the AGN. Quasars are ideal for this since they are detected up to very high redshifts and hence can trace the evolution of actively growing BH mass back into the early Universe (Fan \etal 2001; Fan \etal 2003; Fontanot \etal 2007; Willott \etal 2010) . The BH mass in quasars is calculated using empirical relations derived from optical or UV spectroscopy. In particular, mass-scaling relations between the widths of different broad emission lines and continuum luminosities that have been calibrated against reverberation mapping results \citep{vestergaard_2002, vestergaard_peterson_2006} have allowed BH mass estimates in several large, unobscured (type-1) quasar samples. When translating quasar luminosities into BH masses using the width of broad lines, a similar downsizing is seen in BH mass \citep{vestergaard_osmer_2009, kelly_2010}, suggesting that the most massive BHs ($\Mbh>10^{9}\Msun$) were already in place at $z>2$, whereas the growth of the less massive ones is delayed to lower redshifts. 

An important feature of AGN which should be taken into account is that they exhibit evidence of obscuration at both optical and soft X-ray wavelengths. The obscuration may be linked to the existence of a geometrical torus around the accretion disc whose presence is invoked in the AGN unification scheme \citep{antonucci_1993, urry_padovani_1995} or to intervening dust clouds related to physical processes within the host galaxy \citep{alejo_2005}, such as SF activity (\citealt{ballantyne_2006a}, but also 
\citealt{andy_2009}). As a consequence, a large fraction of AGN could be obscured and thus missing from optical and soft X-ray surveys. Hence, when applying the scaling relations to estimate the mass of accreting BHs, the absence of obscured 
(type-2) quasars from the AGN samples may introduce significant biases into the inferred evolution of BH mass. Only hard X-rays can directly probe the central engine by penetrating the obscuring medium, therefore providing complete and unbiased samples of AGN (\citealt{ueda_2003, lafranca_2005, barger_2005}). However, even in hard X-rays, a population of Compton-thick sources, namely AGN with hydrogen column densities exceeding $N_{\rm H}\simeq10^{24}\rm{cm}^{-2}$, would still be missing \citep{comastri_2004,alexander_2005, alexander_2008, andy_2010}. 

In this paper we present a study of AGN evolution using semi-analytic modelling (see \citealt{baugh_2006} for a review). Our aim is to provide a robust framework for understanding the downsizing of AGN within a self-consistent galaxy formation model. The paper is organized as follows. In Section~2 we present the galaxy formation model upon which we build our AGN model. In Section~3 we study the evolution of BH mass and explore the scaling relations predicted between the BH mass and the mass of the host galaxy mass and dark-matter (DM) halo. In Section~4 we present the essential ingredients of the AGN model and study the evolution of the physical properties which, together with the mass, provide a complete description of accreting BHs. In Section~5 we present our predictions for the evolution of the optical, soft/hard X-ray and bolometric LFs. Finally, in Section~6 we explore the downsizing of AGN in our model. 

\section{The galaxy formation model}

We use the \texttt{GALFORM} semi-analytical galaxy formation code (\citealt{cole_1994, cole_2000}, also \citealt{baugh_2005, bower_2006, font_2008}) and the extension to follow the evolution of BH mass and spin introduced by Fanidakis et~al. (2010). \texttt{GALFORM} simulates the formation and evolution of galaxies and BHs in a hierarchical cosmology by modelling a wide range of physical processes, including gas cooling, AGN heating, SF and supernovae (SN) feedback, chemical evolution, and galaxy mergers.

Our starting point is the \citet{bower_2006} galaxy formation model. This model invokes AGN feedback to suppress the cooling of gas in DM haloes with quasi-static hot atmospheres and has been shown to reproduce many observables, such as galaxy colours, stellar masses and LFs remarkably well. The model adopts a BH growth recipe based on that introduced by \cite{malbon_2007}, and extended by Bower et~al. (2006) and Fanidakis et~al. (2010). During starbursts triggered by a disc instability \citep{efstathiou_1982} or galaxy merger, the BH accretes a fixed fraction, $f_{\mathrm{BH}}$, of the cold gas that is turned into stars in the burst, after taking into account SN feedback and recycling. The value of $f_{\mathrm{BH}}$ is chosen to fit the amplitude of the local observed $\Mbh-M_{\rm bulge}$ relation. In addition to the cold gas channel, quiescent inflows of gas from the hot halo during AGN feedback also contribute to the mass of the BH \citep[see][for the cooling properties of gas in haloes]{white_frenk_1991, cole_2000, croton_2006}. To distinguish between these two accretion channels, we refer to the accretion triggered by disc instabilities or galaxy mergers as the \emph{starburst mode} and the accretion from quasi-hydrostatic hot haloes as the \emph{hot-halo mode}. These correspond to the \emph{quasar} and \emph{radio} modes, respectively, in the terminology used by \citet{croton_2006}. Finally, mergers between BHs, which occur when the galaxies which host the BHs merge, redistribute BH mass and contribute to the build up of the most massive BHs in the universe (see Fanidakis et~al. 2010). 

\texttt{GALFORM} calculates a plethora of properties for each galaxy including disc and bulge sizes, luminosities, colours and metallicities to list but a few. In this analysis, we are primarily interested in the properties that describe the BHs. Specifically, we output the BH mass, $\Mbh$, the amount of gas accreted in a starburst or during the hot-halo mode, $M_{\rm acc}$, the time that has passed since the last burst of SF experienced by the host galaxy and the BH spin (discussed in more detail below). The time since the last burst is necessary to determine the beginning of the active phase of BH growth in the starburst mode.
  
Fanidakis et~al. (2010) extended the model to track the evolution of BH spin, $a$. By predicting the spin we can compute interesting properties such as the efficiency, $\epsilon$, of converting matter into radiation during the accretion of gas onto a BH and the mechanical energy of jets in the \citet{blandford_znajek_1977} and \citet{blandford_payne_1982} mechanisms. To calculate the evolution of $a$ we use the \emph{chaotic} accretion model \citep{king_2008, berti_volonteri_2008, fanidakis_2010}. In this model, during the active phase gas is accreted onto the BH via a series of randomly oriented accretion discs, whose mass is limited by their self-gravity \citep{king_2005}. We choose the chaotic model (over the alternative \emph{prolonged} model in the literature by \citealt{volonteri_2007}) since in this case the predicted BH spin distributions, through their influence on the strength of relativistic jets, reproduce better the population of radio-loud AGN in the local Universe \citep{fanidakis_2010}. A correction to the amount of gas that is accreted is applied in order to account for the fraction of gas that turns into radiation 
during the accretion process. Finally, we note that we do not take into account BH ejection via gravitational-wave recoils during galaxy mergers \citep{merritt_2004}. This omission is not expected to have a significant impact on the evolution of BH mass in our model \citep{libeskind_2006}.

\begin{table}
\caption{Summary of the revised parameter values in the variants of the Bower et al. model considered here.}
\label{models}
\begin{center}
\begin{tabular}{@{}lccc@{}}
\hline
\hline
Study				& ${f_{\rm{Edd}}}^{a}$	& ${\tilde{\epsilon}_{\rm kin}}$$^{b}$		& ${f_{\rm{BH}}}^{c}$	\\       
\hline
Fan10b (this study)		& 0.039				& 0.016			& 0.005			\\
Fan10a (\citealt{fanidakis_2010})	& 0.01				& 0.1			& 0.017			\\      
\hline
\hline
\end{tabular}
\\
\end{center}
{Notes. $^{a}$Fraction of the Eddington luminosity available for heating the gas in the host halo.  $^{b}$Average kinetic efficiency during the hot-halo mode. $^{c}$Ratio of the mass of cold gas accreted onto the BH to the mass of cold gas used to form stars.}
\end{table} 

With this extension to the \citeauthor{bower_2006} model, in \citeauthor{fanidakis_2010} we were able to reproduce the diversity of nuclear activity seen in the local universe.  Furthermore, we demonstrated that the bulk of the phenomenology of AGN can be naturally explained in a $\Lambda$CDM universe by the coeval evolution of galaxies and BHs, coupled by AGN feedback. In this paper we aim to extend the predictive power of the model to high 
redshifts and different wavelengths, to provide a complete and self-consistent framework for the formation and evolution of AGN in a $\Lambda$CDM cosmology. 

For completeness, we now list the model parameters which have an influence over the growth of BH mass:

(i) $f_{\mathrm{BH}}$, the fraction of the mass of stars produced in a starburst, after taking into account SN feedback and the recycling of gas, that is accreted onto the BH during the burst.  

(ii) $f_{\mathrm{Edd}}$, the fraction of the Eddington luminosity of an accreting BH that is available to heat the hot halo during an episode of AGN feedback.

(iii) $\tilde{\epsilon}_{\rm{kin}}$, the average kinetic efficiency of the jet during the hot-halo mode.  

The fiducial model in this paper is denoted as Fan10b. We adopt the values for the above parameters that were used in the Bower et~al. (2006) model, as listed in Table~1 against the Fan10b model. \citeauthor{fanidakis_2010} made some small changes to these parameter values, which are given in the Fan10a entry in Table~1. These changes resulted in a small change in the distribution of accretion rates in the hot-halo mode. By reverting to the original parameter values used in the Bower et~al. model, we note that there is a small tail of objects accreting in the hot-halo mode, for which the accretion rate is higher than that typically associated with advection dominated accretion via a thick disc (see Section~4.1 for further discussion). 

\begin{figure}
\center
\includegraphics[scale=0.42]{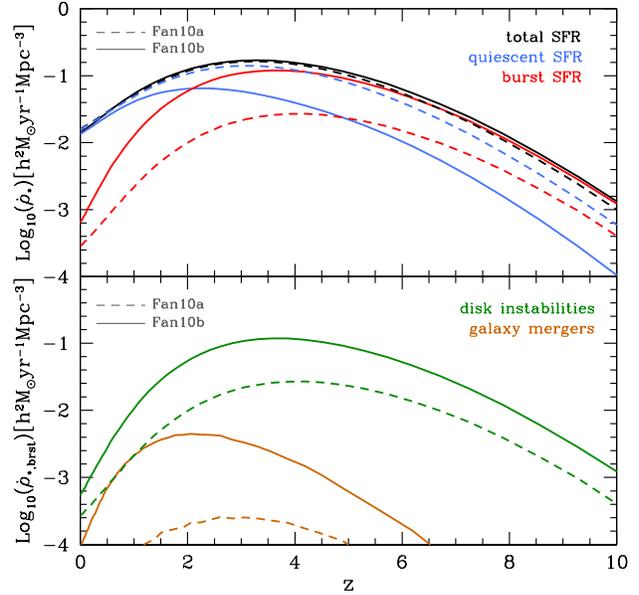}
\caption{Top: The cosmic history of the total SFR density (black lines) in the Fan10b (solid line) and Fan10a (dashed line) models. Also shown is the contribution from each SF activity mode, namely burst and quiescent SF (red and blue respectively), to the total SFR density. Bottom: The contribution of disc instabilities (green lines) and galaxy mergers (orange lines) to the cosmic history of SFR density in the burst SF activity mode for the Fan10b (solid lines) and Fan10a (dashed lines) models.} 
\label{sfr}
\end{figure}

A further difference between our fiducial model, Fan10b, and the Fan10a model is the use of an improved SF law. Following Lagos et~al. (2010), who implemented a range of empirical and theoretical SF laws into \texttt{GALFORM}, we use the SF law of Blitz \& Rosolowsky (2006, hereafter BR06). Lagos et~al. found that this SF law in particular improved the agreement between the model predictions and the observations for the mass function of atomic hydrogen in galaxies and the hydrogen mass to luminosity ratio. The BR06 model has the attraction that it is more physical than the previous parametric SF law used in \citeauthor{bower_2006}, and agrees with observations of the surface density of gas and SF in galaxies. The BR06 law distinguishes between molecular and atomic hydrogen, with only the molecular hydrogen taking part in SF. The fraction of hydrogen in molecular form depends on the pressure within the galactic disc, which in turn is derived from the mass of gas and stars and the radius of the disc; these quantities are predicted by \texttt{GALFORM}. The BR06 SF law contains no free parameters once it has been calibrated against observations. We adopt a value for the normalization of the surface density of SF that is two thirds of the value used by Lagos et~al., but which is still within the observational uncertainty. We do this to improve the match to the observed quasar luminosity function. 

\begin{figure}
\center
\includegraphics[scale=0.42]{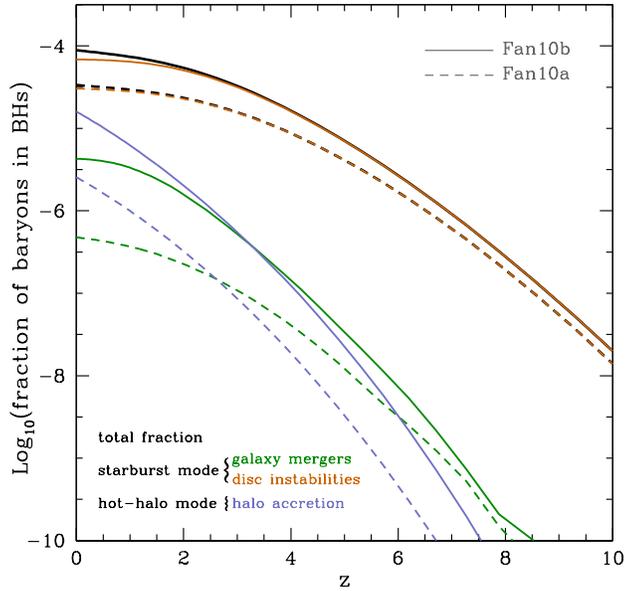}
\caption{The fraction of baryons locked in BHs as a function of redshift (black lines) for the Fan10b (solid lines) and Fan10a (dashed lines) models. The plot also shows the contribution of each accretion channel, namely disc instabilities (orange), galaxy mergers (green) and quasi-hydrostatic halo accretion (blue), to the total fraction of baryons in BHs.} 
\label{baryons_SMBHs}
\end{figure}

As shown by Lagos et~al., the adoption of the BR06 SF law changes the SF history predicted by the model. This is due to a much weaker dependence of the effective SF timescale on redshift with the new SF law, compared with that displayed by the Bower et~al. model. This leads to the build-up of larger gas reservoirs in discs at high redshift, resulting in more SF in bursts. This is shown in the top panel of Fig.~\ref{sfr}, which compares the predictions of the Fan10b (solid lines) and Fan10a (dashed lines) models for the total SFR density and the SFR density in the quiescent and burst SF modes. Fig.~\ref{sfr} shows that the individual SF modes change substantially on using the BR06 SF law. The quiescent SF mode in the Fan10b model is suppressed by almost an order of magnitude above $z\sim3$ compared to the Fan10a model. This has an impact on the build up of BH mass, by changing the amount of mass brought in through the starburst mode. The enhancement of the burst mode is further demonstrated in the bottom panel of Fig.~\ref{sfr}, where we show the SFR density history in bursts distinguishing between those triggered by disc instabilities and galaxy mergers for the Fan10a and Fan10b models. Both channels show a significant increase in SFR density in the Fan10b model. Since the BHs in our model grow in bursts of SF following disc instabilities and galaxy mergers we expect this enhancement to have a significant impact on the evolution of BH mass. 

\begin{figure*}
\center
\includegraphics[scale=0.67]{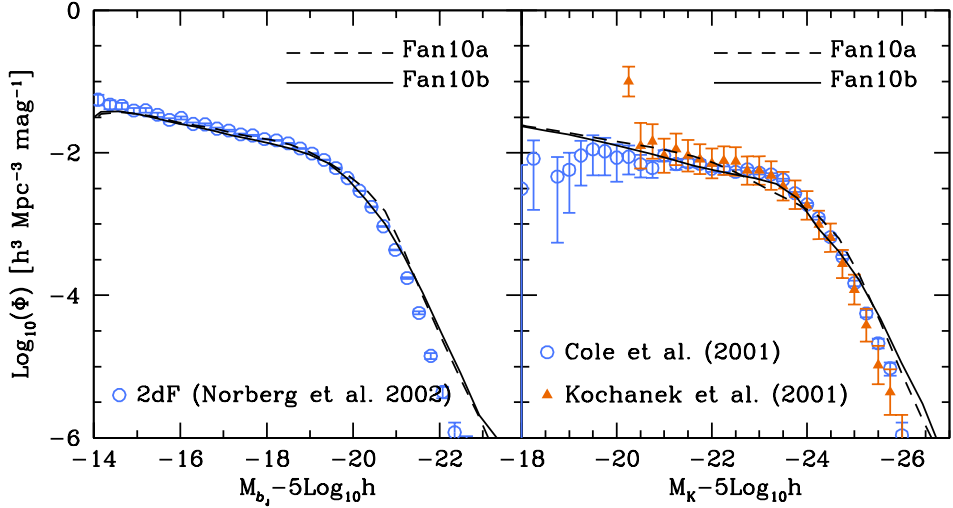}
\caption{The LF of galaxies in the local Universe. The left panel compares the predictions of the Fan10b (solid line) and Fan10a (dashed line) models for the $\bj$-band LF with the observational determination from the 2dF galaxy redshift survey by Norberg et al. (2002). Similarly, the right panel shows the predictions of the models (line style as before) for the K-band LF and compared to the observational determinations by Cole et al. (2001) and Kochanek et al. (2001). The theoretical predictions from both models include dust extinction.} 
\label{galaxy_lf}
\end{figure*}

To evaluate the change in the BH mass when we use the BR06 SF law, we plot in Fig.~\ref{baryons_SMBHs} the fraction of baryons locked up in BHs as a function of redshift. In this plot we have also distinguished between the different accretion channels: disc instabilities, galaxy mergers and accretion from hot gas haloes in quasi-hydrostatic equilibrium. We note that in this plot it becomes clear that the growth of BHs is dominated by accretion of cold gas during disc instabilities. Therefore, secular processes in galaxies are responsible for building most of the BH mass, while galaxy mergers become an important channel only when they occur between galaxies of similar mass (i.e. major mergers). 

The contribution of the starburst mode to the build up of BH mass increases significantly in the Fan10b model compared with the Fan10a model. Both the disc instability and galaxy merger channels are enhanced by a significant factor, resulting in different evolution of the total BH mass. The Fan10b model also predicts different hot-halo mode accretion. This can be attributed to the different values of the $\tilde{\epsilon}_{\rm{kin}}$ and $f_{\mathrm{Edd}}$ parameters used in Fan10a (note that halo accretion is completely independent of SF in the galactic discs). It will become evident in later sections that all of the aforementioned changes induced to the BH mass are essential for providing a good fit to the evolution of the AGN LF. 

Finally we note that the total SFR density history in Fig.~\ref{sfr} remains unchanged when using the BR06 SF law, as reported by Lagos et~al. This implies that properties such as stellar masses and galaxy luminosities are fairly insensitive to the choice of SF law. Indeed, \citeauthor{lagos_2010} show that the predictions for the $\bj$ and $K$-band LFs on using the BR06 SF law are very similar to those of the original \citeauthor{bower_2006} model and reproduce the observations very closely. Since we have changed the normalization of the SF surface density, we recheck that our model still reproduces the observed LF of local galaxies, as shown in Fig.~\ref{galaxy_lf}. Hence, we are confident that we are a building an AGN model within a realistic galaxy formation model. 

\section{The evolution of BH mass}\label{sec:BH mass evolution}
BH demographics have been the topic of many studies in the past decade mainly because of the tight correlations between the properties of BHs and their host stellar spheroids. These correlations take various forms, relating, for example, the mass of the BH to the mass of the galactic bulge \citep[the $\Mbh-M_{\rm Bulge}$ relation:][]{magorrian_1998, mclure_2002, marconi_hunt_2003, haring_rix_2004}, or to the stellar velocity dispersion \citep[the $\Mbh-\sigma$ relation:][]{ferrarese_merritt_2000, gebhardt_2000, tremaine_2002}. These remarkable and unexpected correlations suggest a natural link between the BH evolution and the formation history of galaxies. The manifestation of this link could be associated with AGN activity triggered during the build up of BHs.
\begin{table}
\caption{Summary of the local BH mass densities found in this and previous studies (assuming $h=0.7$). Densities are shown for BHs in disc (S,S0) and elliptical (E) galaxies, and for the global BH population in our model (tot).}
\label{rho_bh}
\begin{tabular}{@{}lccc@{}}
\hline
\hline
Study 				& $\rbh(\mathrm{S,S0})^{1}$ & $\rbh(\mathrm{E})^{2}$ & $\rbh(\mathrm{tot})^{3}$ \\
\hline
This study 				& 0.72 				& 4.70 				& 5.42 \\
\citet{graham_2007}		& $0.95^{+0.49}_{-0.49}$ 		& $3.46^{+1.16}_{-1.16}$  		& $4.41^{+1.67}_{-1.67}$ \\
\citet{shankar_2004}       	& $1.1^{+0.5}_{-0.5}$ 	& $3.1^{+0.9}_{-0.8}$	& $4.2^{+1.1}_{-1.1}$  \\
\citet{marconi_2004}      	& $1.3$ 				& 3.3	                      	& $4.6^{+1.9}_{-1.4}$ \\
\citet{fukugita_2004}  	& $1.7^{+1.7}_{-0.8}$  	& $3.4^{+3.4}_{-1.7}$  	& $5.1^{+3.8}_{-1.9}$ \\
\hline
\hline
\end{tabular}
\\
{Notes. $^{1,2,3}$In units of $10^{5}\Msun\rm{Mpc}^{-3}$.}
\end{table} 

The observed $\Mbh-M_{\rm Bulge}$ relation can be used to estimate the mass of a BH. Several authors have utilised this technique to estimate the BH mass function (BHMF) in the local Universe \citep{yu_tremaine_2002, marconi_2004, shankar_2004}, using scaling relations such as $\Mbh-\sigma_{*}$ and $\Mbh-L_{\mathrm{Bulge}}$. Based on these MFs, and others inferred by independent studies, the total BH mass density in the local Universe has been estimated to be in the range $\rbh=(4.2-5.1)\times10^{5}\Msun\mathrm{Mpc}^{-3}$ (see Table~\ref{rho_bh} for a list of local BH mass density estimates). Note that, when correcting for unknown dependencies on $h$, the different estimates from the literature suggest a BH mass density consistent with $4.4-5.9\times10^{5}\Msun\rm{Mpc}^{-3}$ \citep{graham_2007}.  

\begin{figure*}
\center
\includegraphics[scale=0.33]{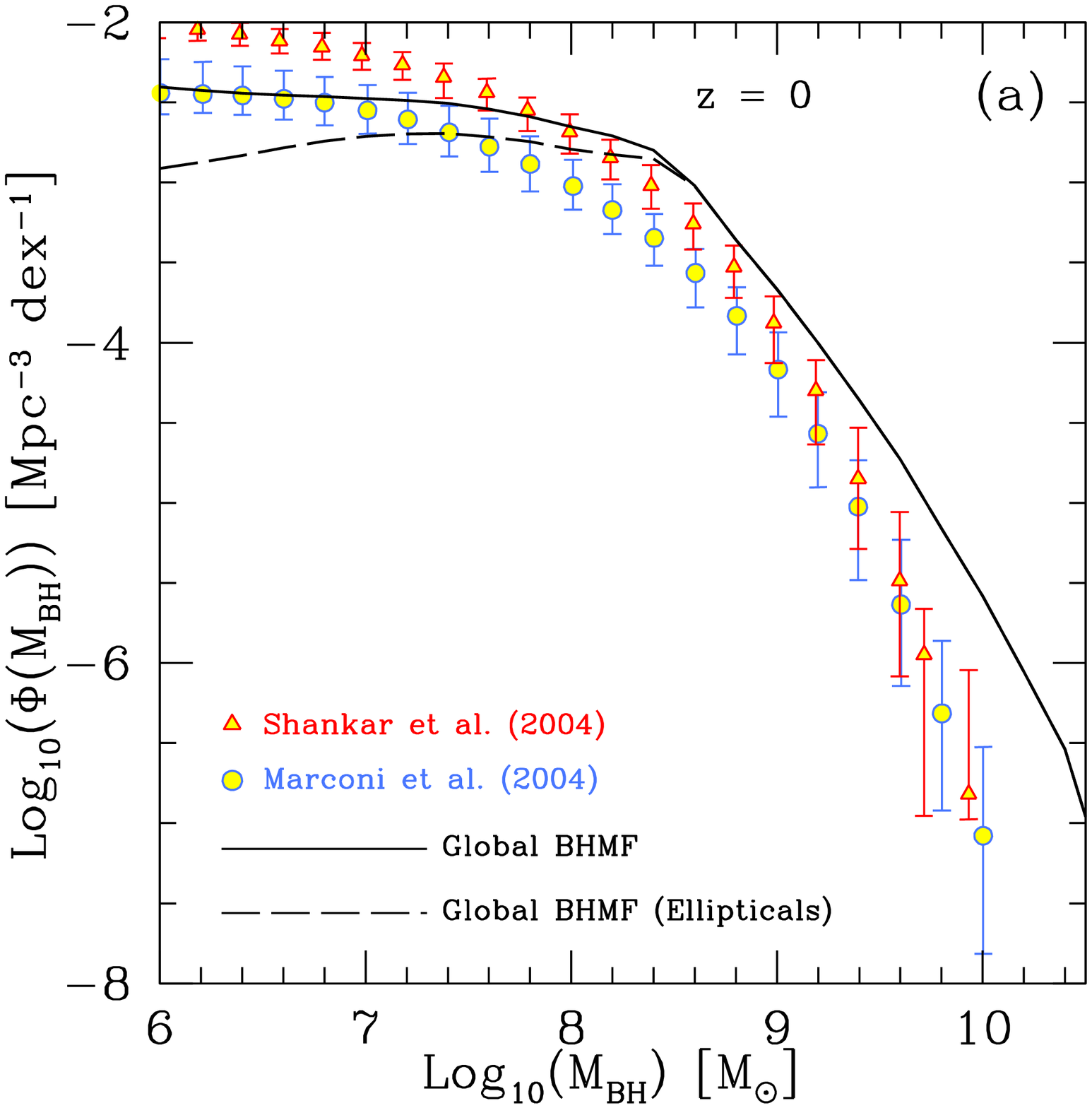}
\includegraphics[scale=0.33]{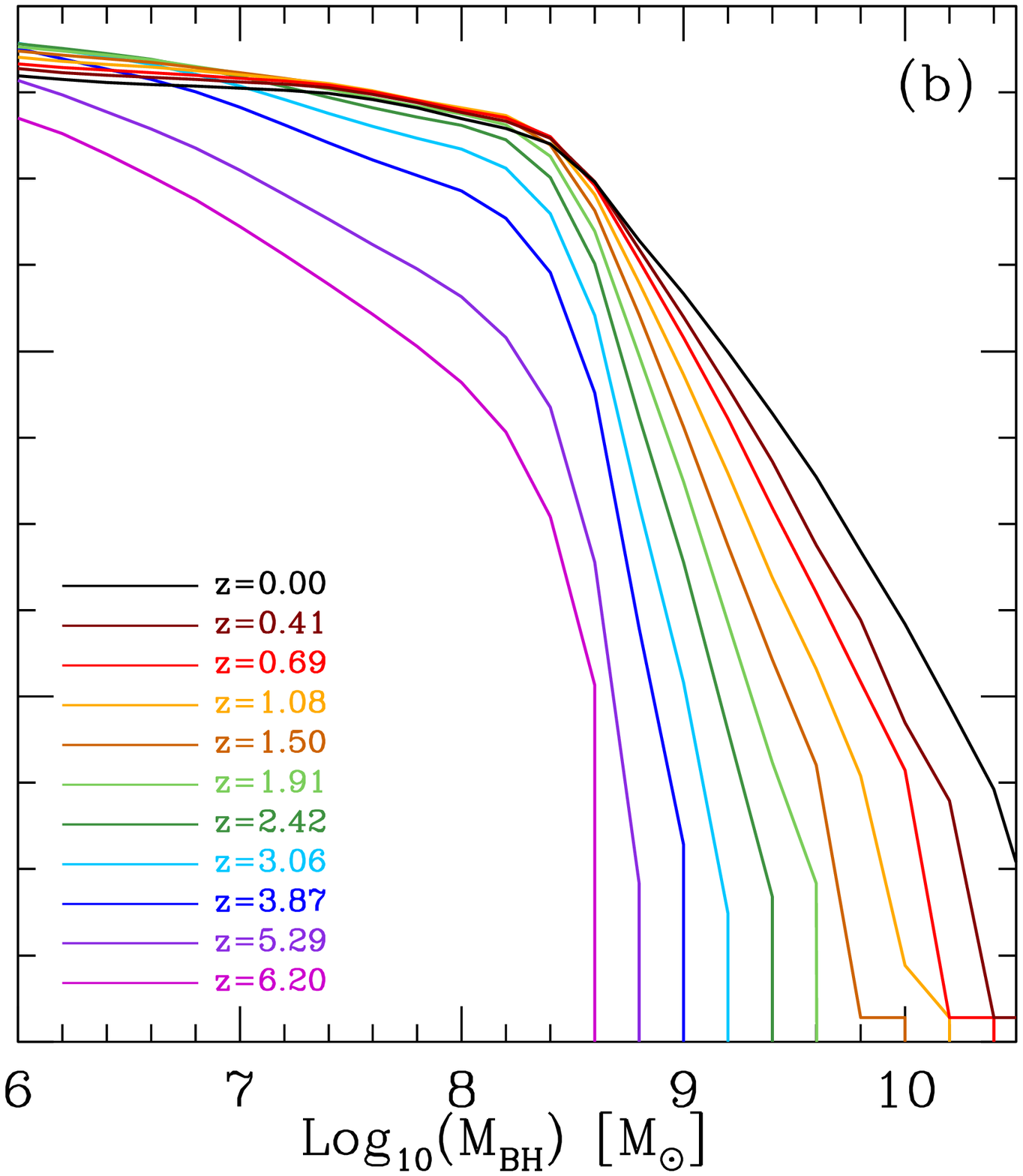}
\includegraphics[scale=0.33]{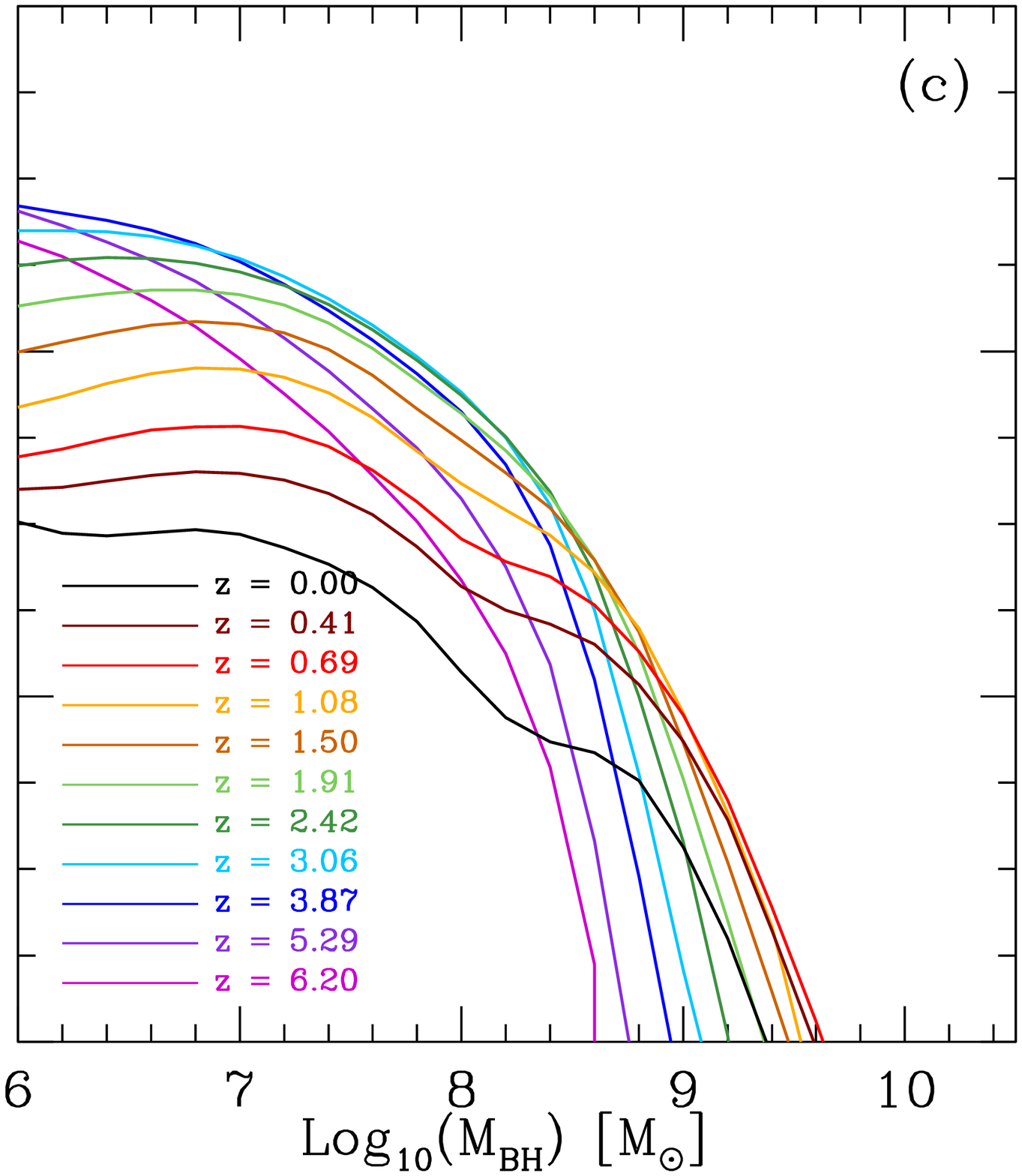}
\caption{a) The global BHMF at $z=0$ predicted by our model (black solid line). The observationally determined MFs are taken from \citet[][red filled triangles]{marconi_2004} and \citet[][blue filled circles]{shankar_2004}. The error bars correspond to $\pm1\sigma$ uncertainties. b) The evolution of the global BHMF with redshift. c) The MF of actively growing BHs (BH accreting in the thin-disc regime) calculated at different redshifts as indicated in the legend.} 
\label{bhmf}
\end{figure*}

The global BHMF predicted by the model for all BHs at $z=0$ is shown in Fig.~\ref{bhmf}a. The predicted MF is almost constant in the mass range $10^6-10^8\Msun$. For higher masses the MF decreases steeply with increasing mass. Our predictions are compared to the local MF estimated by \citet{marconi_2004} and \citet{shankar_2004} using the BH mass scaling relations in local galaxies. The predicted and observed MFs are in good agreement for low mass BHs and disagrees moderately for the $10^9-10^{10}\Msun$ BHs. The overall BH mass density at $z=0$ predicted by our model amounts to $\rbh=5.42\times10^5\Msun\mathrm{Mpc}^{-3}$ and is consistent with that estimated by \citeauthor{marconi_2004} ($\rbh=4.6^{+1.9}_{-1.4}\times10^5\Msun\mathrm{Mpc}^{-3}$) and other authors \citep[see Table~\ref{rho_bh}]{shankar_2004, graham_2007, yu_lu_2008}. Note that, when calculating the BH mass density we do not take into account mass loses due to gravitational wave emission \citep{menou_2004}. 

In Fig.~\ref{bhmf}a we also show the contribution to the MF from BHs hosted by elliptical galaxies. We classify as ellipticals all the galaxies whose bulge luminosity contributes more than 60\% to their total $\bj$-band luminosity \citep{cole_2000, parry_2009}. The rest of the galaxies are classified as spiral or S0 galaxies. As expected, the BHs in the elliptical galaxies dominate the high mass end of the MF. Hence, the most massive BHs inhabit the centres of the most massive systems in our simulations. By contrast, spiral and S0 galaxies contribute only to the low-mass end of the MF. 

The growth of BHs in spiral and elliptical galaxies is driven in principle by different channels. In spiral galaxies, BHs grow mainly via accretion during the starburst mode whereas the massive BHs in ellipticals grow via accretion during the hot-halo mode or BH-BH mergers. Evidently, the evolution of the BHMF should be shaped by these growth channels. This is illustrated in Fig.~\ref{bhmf}b, where we show the evolution of the global BHMF in the redshift range $z=0-6.2$. The build up of the BHs populating the low-mass end takes place at high redshifts and is complete at $z\sim2$. Since $z\sim2$ the amplitude of the BHMF remains almost unchanged indicating that $10^{6}-10^{8}\Msun$ BHs in spiral galaxies today were already in place by $z\sim2$, which follows from the spheroid being older than the disc. The growth of these BHs is dominated by accretion during the starburst mode. 

The fact that the space density of this population does not evolve since $z\sim2$ suggests that the starburst mode becomes less significant in the low-redshift universe. Nonetheless, the build up of the BH mass continues below $z\sim2$, where we witness the build up of the most massive BHs in the universe. These BHs grow mainly during the hot-halo mode or via mergers with other BHs \citep[see][]{fanidakis_2010}. The different physics governing these channels results in a different slope for the high-mass end of the BHMF giving rise to a strong break at $\sim5\times10^{8}\Msun$. Yet, the build up of these BHs is not very efficient. This becomes clearer when we consider the evolution of the BH mass density. For instance, in the redshift range $z=6.2-1.9$ the BH mass density increases by a factor of $\sim25$, as a result of the intense accretion during the starburst mode in the high redshift universe. By contrast, from $z=1.9$ until $z=0$ the density only doubles, mainly because the $10^{9}\Msun$ BHs become more numerous.

To gain more insight into the rapidly evolving population of accreting BHs we consider the evolution of actively growing BHs, namely those BHs that accrete at relatively high rates (greater that $0.1\%$ of their Eddington accretion rate, see Section \ref{sec:Disc luminosities}). In Fig.~\ref{bhmf}c we show the MF of actively growing BHs in our model and its evolution with redshift. The most striking characteristic of the MF is the dramatic change in shape and normalization with redshift. At $z=0$ the MF has a maximum at $\sim10^{7}\Msun$ below which the accreting BHs have an almost constant space density \citep[see also results from SDSS][]{heckman_2004}. A significant bend in the slope located at $\sim10^{8}\Msun$ is seen, a feature that remains evident up to $z\sim2-3$. At higher redshifts, the space density of BHs drops substantially, establishing that accreting BHs with masses $\gtrsim10^{9}\Msun$ are very rare objects. These BHs are accreting at relatively low rates and therefore power faint-AGN activity (see discussion in Section~\ref{sec:The Lbol-Mbh correlation}).

As illustrated by Fig.~\ref{bhmf}c, the low-mass end of the MF evolves differently with redshift compared to the high-mass end. Its amplitude increases significantly with increasing redshift, suggesting that the space density of low-mass accreting BHs was higher at earlier epochs. This evolution is consistent up to $z\sim2.5$ above which the amplitude of the low-mass end starts to decline modestly. The high-mass end shows a similar increase towards higher amplitudes with increasing redshift, however, the decline appears earlier, almost at $z\sim1$. This is because the high-mass end forms much later than the low-mass end in our model. 

The change of the BHMF with redshift illustrated in Fig.~\ref{bhmf}c indicates an evolutionary scenario for the actively growing BH population in our model which mirrors the hierarchical growth of their host galaxies. At high redshifts the accretion activity is dominated by $10^6\Msun$ BHs. Accretion onto these BHs results in the fast build up of the $10^7-10^8\Msun$ BHs at $z\sim4-6$, as implied by the increase in the space density of these BHs in Fig.~\ref{bhmf}c. From $z\sim4$ the accretion activity, gradually shifts to the $10^7-10^8\Msun$ BH population. This is mainly because more higher mass galactic systems are now in place and thus disc instabilities and galaxy mergers contribute to the growth of higher mass BHs compared to earlier epochs. Accretion onto these BHs leads to the fast build up of the $>10^8\Msun$ BH population, which will later be promoted via accretion during the hot-halo mode and BH-BH mergers to the most massive BHs in our model.

\begin{figure*}
\center
\includegraphics[scale=0.43]{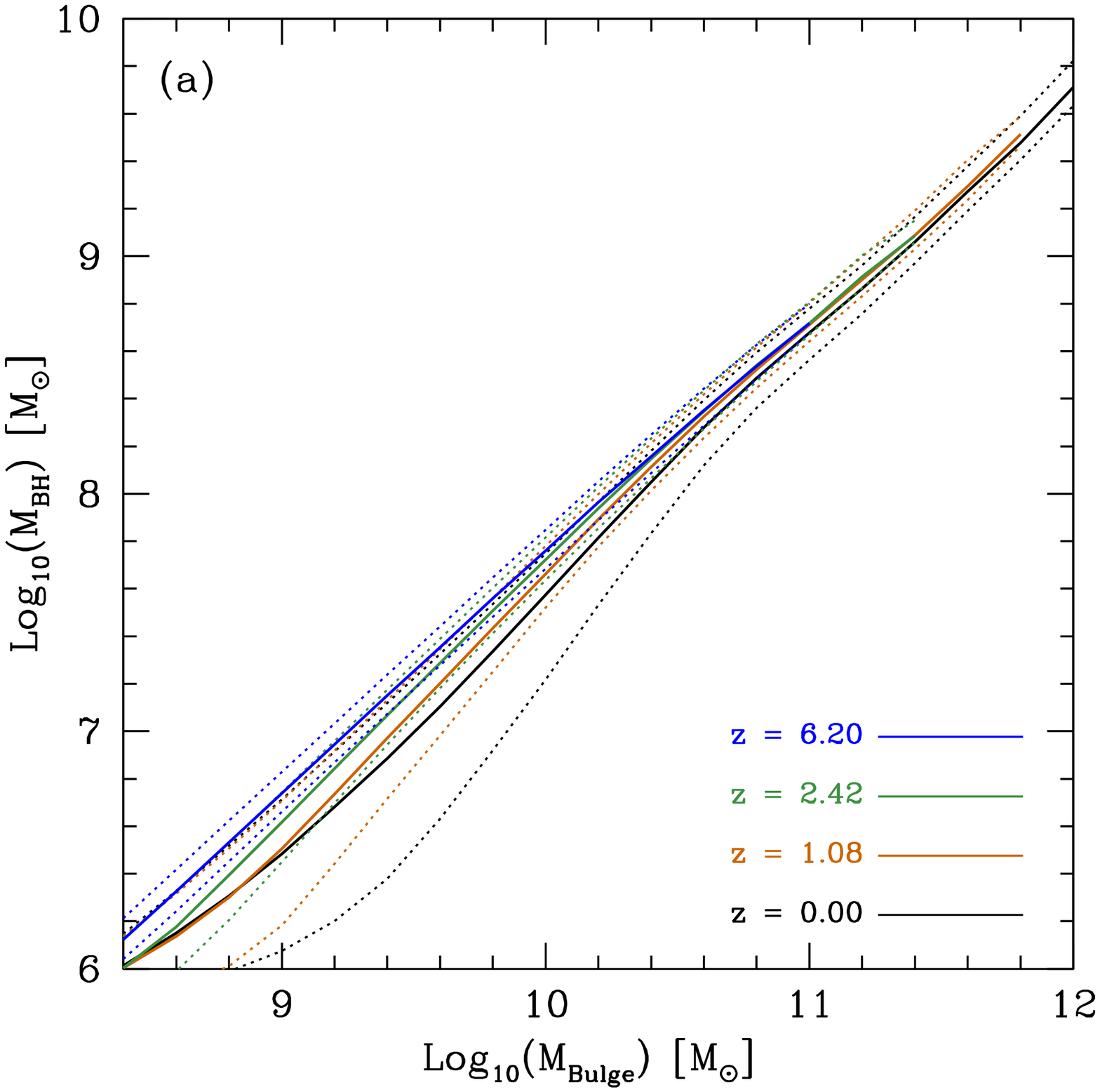}
\includegraphics[scale=0.43]{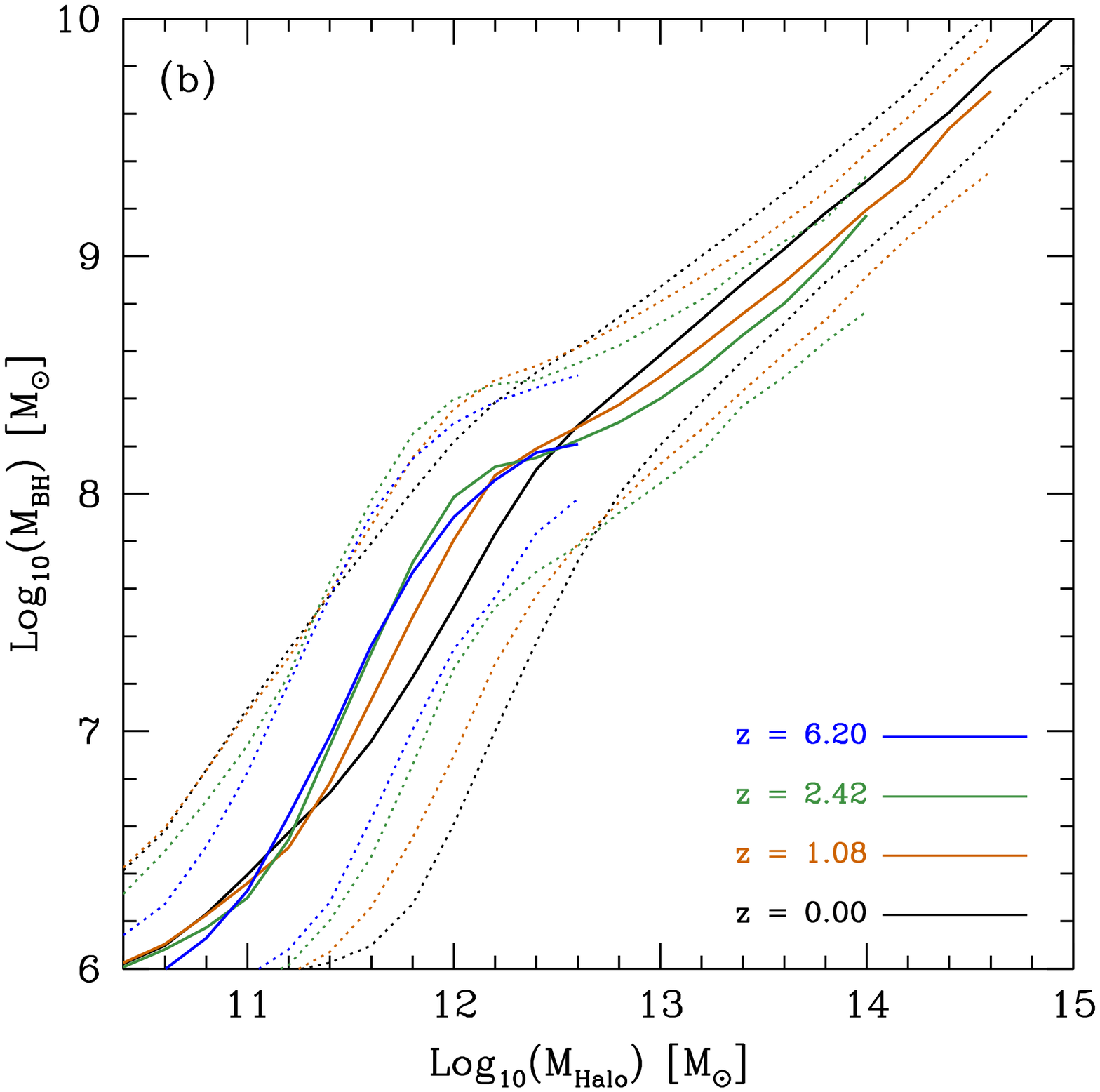}
\caption{The median of the $\Mbulge-\Mbh$ (a) and $M_{\rm{Halo}}-\Mbh$ (b) distributions predicted by our model (solid lines). Predictions are shown for $z=0,1.08,2.42$ and $6.20$ as indicated by the legend. The dotted lines indicate the $10-90$ percentile ranges of the distributions.}  
\label{scal_relations}
\end{figure*}

Eventually, the different physical processes that drive the evolution of BH mass result in a tight correlation between the BH and host galaxy mass, as shown in Fig.~\ref{scal_relations}a. In this figure, we plot the median of the $\log\Mbh-\log\Mbulge$ distribution (solid lines) and the associated $10-90$ percentiles (dotted lines) at $z=0,1.08,2.42$ and $6.2$. Surprisingly, a well defined $\Mbh-\Mbulge$ relation is established already at $z=6.2$, merely as a consequence correlation of the accretion process with the SF that shapes the mass of the host bulges. Even when the growth of BHs is dominated by other processes, which are only indirectly linked to the SF, the tight correlation remains. For example, the most massive BHs correlate with the mass of their host bulge because feedback energy when gas is accreted onto them regulates the cooling flows that supply gas for SF and therefore the change of mass of the bulge.

In a hierarchical universe the most massive galactic bulges are usually found in the most massive DM haloes. Since the galaxies with the most massive bulges (namely the passive ellipticals) host the most massive BHs and those with less massive bulges (namely the spiral and S0 galaxies) host the low-mass BHs, it is natural to assume that there must be a similar hierarchy in the halo environments where these BHs can be found. Indeed, when plotting the relation between the BH mass and the mass of the host halo we find a tight correlation. This is illustrated in Fig.~\ref{scal_relations}b where we show the median of the $\log M_{\rm{Halo}}-\log \Mbh$ distribution (solid lines) and its $10-90$ percentiles (dotted lines) for different redshifts. Remarkably, the physical processes that shape the $M_{\rm{Bulge}}-\Mbh$ relation in our model give rise to a well defined link between these two quantities at all redshifts. The $M_{\rm{Halo}}-\Mbh$ relation can be parametrized by a power law with a very steep slope that becomes shallower above $M_{\rm{Halo}}\simeq10^{12}-10^{13}\Msun$. 

The different slopes indicate the different efficiency with which BHs grow in haloes of different mass. Low mass haloes are very efficient in growing BHs since in these environments the gas cooling is not suppressed. As a consequence, BHs double their mass remarkably quickly resulting in a steep slope for the $\log M_{\rm{Halo}}-\log \Mbh$ relation. The fast BH mass build up slows down when hot gas in the host haloes enters the quasi-hydrostatic regime. In this case, AGN feedback suppresses the cooling flows that provide fresh cold gas to the galaxies and thus accretion during the starburst mode is reduced. In the quasi-hydrostatic regime gas accretion during the hot-halo mode and mergers dominate the BH mass build up. The hot-halo mode is, however, characterized by low accretion rates (see Section~\ref{sec:The Ledd parameter}). BH mergers rather than adding new baryons to the BHs, only redistribute the BH mass. Therefore, the BH mass build up slows down establishing haloes of $M_{\rm{Halo}}\sim10^{13}-10^{15}\Msun$ as environments where BH growth is not very efficient. 

Given the efficiencies that characterize the two different regimes in Fig.~\ref{scal_relations}b we expect to find the brightest AGN in the $10^{11}-10^{13}\Msun$ haloes. This perhaps is in contrast with the common expectation that the brightest quasars should be found in the most massive haloes. 

\section{BH spins, accretion efficiencies and disc luminosities}\label{sec:Disc luminosities}

AGN are unambiguously the most powerful astrophysical sources in the Universe. The ability of an accretion disc to produce electromagnetic radiation is attributed to gravity yet its radiative efficiency is controlled primarily by the properties of the gas. When the gas settles itself into a thin, cool, optically-thick accretion disc \citep{shakura_1973}, the efficiency of converting matter into radiation can reach $32\%$ (Novikov \& Thorne 1973). When the flow is characterized by very high, Eddington accretion rates, such extremely luminous discs can power $10^{47}-10^{48}\ergsec$ quasars.  In the next sections we describe how we model the physics of accretion flows and present predictions for the most fundamental properties characterising the accreting BH systems in our model.

\subsection{Calculation of the disc luminosity}\label{sec:calculation of the disc luminosity}
The first important property that we can calculate in our model is the physical accretion rate, $\dot{M}$, onto a BH. This is defined as 
\begin{equation}
\dot{M}=\frac{M_{\mathrm{acc}}}{t_{\mathrm{acc}}},
\end{equation}
where $M_{\mathrm{acc}}$ is the total accreted mass and $t_{\mathrm{acc}}$ is the accretion timescale. The accretion timescale is assumed to be directly linked to the dynamical timescale of the host bulge, $t_{\mathrm{Bulge}}$, through the relation
\begin{equation} 
t_{\mathrm{acc}}=\fq t_{\mathrm{Bulge}}.
\label{acc_timescale}
\end{equation}
$\fq$ is a free parameter and its value is fixed to 10. This is determined by fitting the predictions of our model for the quasar LF to the observations (see Section~\ref{sec:The optical LF}). 

A second important quantity associated with every BH is the Eddington luminosity,
\begin{equation}
\Ledd=\frac{4\pi G\Mbh m_{\mathrm p} c}{\kappa}=1.4\times10^{46}\left(\frac{\Mbh}{10^8~\Msun}\right)~\mathrm{erg~s}^{-1},
\end{equation}
where $\kappa\sim0.3$~cm$^2$g$^{-1}$ is the electron scattering opacity and $m_{\mathrm p}$ is the proton mass. The Eddington luminosity has an associated accretion rate which is expressed as 
\begin{equation}
\Medd =\Ledd/\epsilon c^2,
\end{equation}
where $\epsilon$ is the accretion efficiency and $c$ is the speed of light. This is the accretion rate at which the black hole radiates at the Eddington luminosity. The accretion efficiency is assumed to be determined by the spin of the BH (Novikov \& Thorne 1973) and its value is calculated as in \citet{fanidakis_2010}. It is convenient in our analysis to express the physical accretion rate in units of the Eddington accretion rate, $\dot{m}\equiv\dot{M}/\Medd$, in order to introduce a dependence on the BH mass. 

The accretion rate has a dramatic impact on the geometry and radiative properties of the accretion disc \citep[see][]{done_2007}. In our model, we consider two distinct accretion modes separated at a rate of $1\%$ of $\Medd$. In the first state, for $\m\geqslant0.01$, we assume that the gas forms an accretion disc whose physics is adequately described by the radiatively efficient thin-disc model of \citet{shakura_1973}. The bolometric luminosity of a thin disc, $\Lbol$, is linked to $\dot{M}$ through the standard expression
\begin{equation}
\Lbol = \epsilon\dot{M}c^2.
\end{equation}
When the accretion becomes substantially super-Eddington ($\Lbol\geqslant\eta\Ledd$), the bolometric luminosity is limited to $\eta[1+\ln(\dot{m}/\eta)]L_{\mathrm{Edd}}$ \citep{shakura_1973}, where $\eta$ is an ad hoc parameter equal to $2$ that allows a better modelling of bright end of the LF (see Section~\ref{sec:The evolution of LFs}). However, we do not restrict the accretion rate if the flow becomes super-Eddington.

The second accretion state, with $\dot{m}<0.01$, is associated with sub-Eddington accretion flows with very low density. For such low accretion rates, the gas flow is unable to cool efficiently since radiative cooling does not balance the energy generated by viscosity. Thus, the viscous energy is trapped in the gas as entropy and ultimately advected into the hole. This type of accretion is known as an advection dominated accretion flow \citep[ADAF,][]{rees_1982, narayan_1994, abramowicz_1995}. ADAFs have a number of distinct properties, some of which will be essential for the analysis in later sections (see Sections 5 and 6). For example, for an ADAF around a BH, only a fraction of the standard accretion luminosity, $L=\epsilon\dot{M}c^2$, is emitted as radiation. The remainder of the viscously dissipated energy is stored in the gas as entropy, resulting in hot flows with almost virial temperatures. We note that, as shown by \citet{ichimaru_1977}, the ions and the electrons in an ADAF are not thermally coupled and, thus, reach different temperatures. This two-temperature virialized plasma flow is optically thin and, for high viscosity parameters ($\alpha\sim0.1-0.3$), it acquires a quasi-spherical geometry around the BH, which resembles spherical Bondi accretion. However, the accretion is entirely due to dissipation via viscous forces rather than gravity. The bolometric luminosity of the flow in this case is equal to the luminosity emitted by the various cooling processes. For $\dot{m}\lesssim10^{-3}\alpha^2$ it is only due to Comptonisation, whereas  for $\dot{m}\gtrsim10^{-3}\alpha^2$ the cooling is split between Compton and synchrotron emission.

As the accretion rate increases, the emission of the energy produced by the viscous processes in the gas becomes more efficient. Above some critical accretion rate, $\dot{m}_c$, the radiative efficiency of the gas is so high that the flow cools down to a thin disc. The critical accretion is independent of the BH mass but depends strongly on the viscosity, $\dot{m}_{c}\simeq1.3\alpha^2$. Taking $\dot{m}_{c}=0.01$ implicitly fixes the value of $\alpha$ to be $0.087$ for all the ADAFs in our model. The luminosity of the flow for the two different regimes is then given by \citet{mahadevan_1997},
\begin{eqnarray}
L_{\mathrm{bol,ADAF}}=\left\{\begin{array}{@{}ll@{}}\epsilon\dot{M}c^2 \negthinspace\left[0.4\dot{m}\beta/\alpha^2\right], \negthinspace & \negthinspace\dot{m}>7.5\times10^{-6} \\ 
\epsilon\dot{M}c^2 \negthinspace\left[4 \negthinspace\times \negthinspace 10^{-4}(1 \negthinspace-\negthinspace\beta)\right], \negthinspace &\negthinspace \dot{m}\lesssim7.5\times10^{-6}\end{array}\right.
\label{disc_bol_adaf}
\end{eqnarray}
where $\beta$ is related to the Shakura-Sunyaev viscosity parameter $\alpha$ through the relation $\alpha\approx0.55(1-\beta)$ \citep{hawley_2005}. The expressions in the square brackets in Eq.~\ref{disc_bol_adaf} shows how much less efficient the cooling is in an ADAF compared to the standard efficiency $\epsilon$ of a thin disc. For example, at $\dot{m}=0.01$ an ADAF is characterized by an accretion efficiency of $0.44\epsilon$, less than approximately half as luminous as a thin disc.

\subsection{BH spins and accretion efficiencies}

\begin{figure}
\center
\includegraphics[scale=0.43]{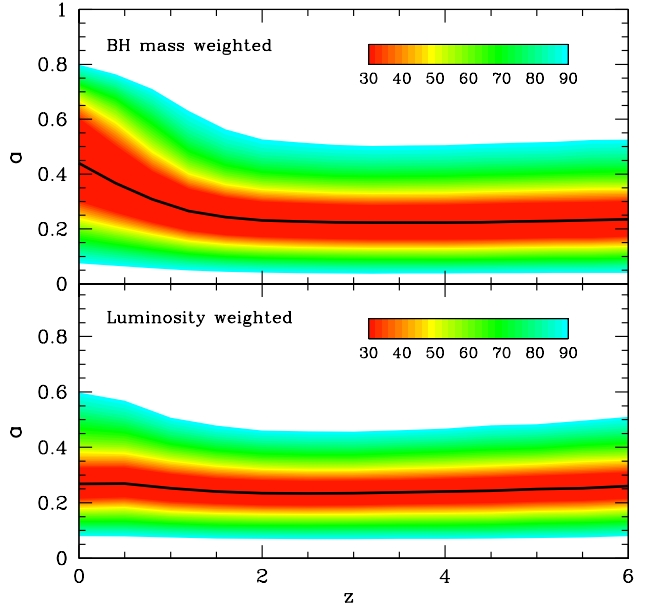}
\caption{Top: The median BH spin weighted by BH mass as a function of redshift for BHs in our model (solid black line). Bottom: The median BH spin weighted by disc luminosity for all AGN accreting in the thin-disc regime (solid black line). The values of the different percentiles (shaded regions) are indicated in both cases by the colour bars.} 
\label{spin_evolution}
\end{figure}

In \citet{fanidakis_2010} we describe in detail how we model the evolution of BH spin. In brief, the spins of BHs change during gas accretion (whenever a starburst is triggered by disc instabilities or galaxy mergers in the starburst mode or during the cooling of gas from the hydrostatic halo in the hot-halo mode) and mergers with other BHs. Fig.~\ref{spin_evolution} shows the evolution with redshift of the median of the BH spin distribution at a given redshift for two different cases. Firstly, we show the median spin weighted by BH mass (upper panel) for all the BHs with $\Mbh>10^6\Msun$. The median shows an approximately constant trend from $z=6$ to $z=2$ and has a well defined value of $\sim0.25$.  In this redshift range BHs grow predominantly during disc instabilities (see Fig.~\ref{baryons_SMBHs}) and therefore the evolution of BH spin is governed by accretion. As shown by \citeauthor{fanidakis_2010}, accretion of gas results in low spins with a typical value of $a\sim0.2$ (under the assumption that the gas is fed chaotically onto the BH). Hence, as indicated also by the different percentiles in the plot (shaded regions), the bulk of BHs acquire low spins. 

\begin{figure*}
\center
\includegraphics[scale=0.229]{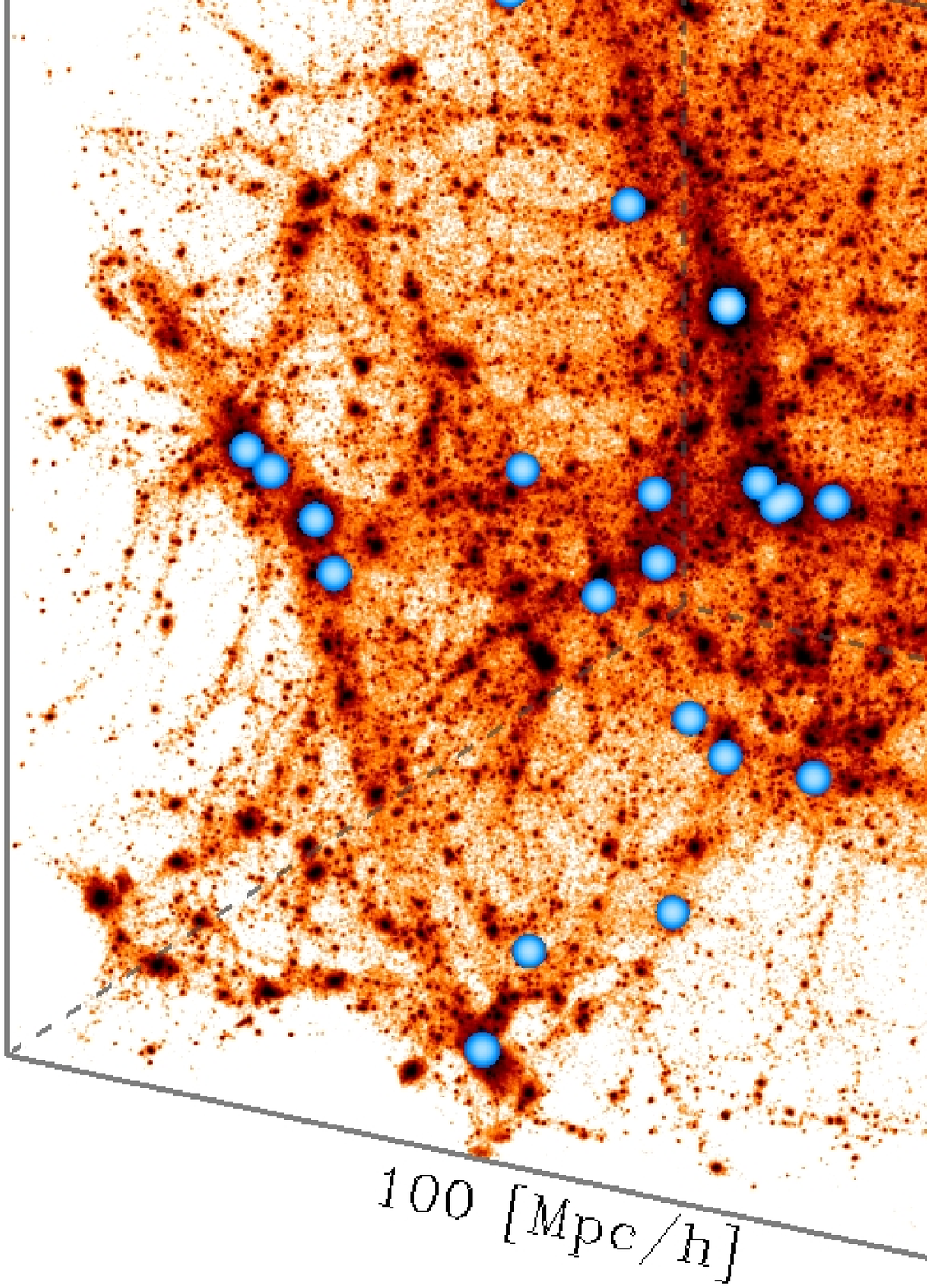}
\includegraphics[scale=0.229]{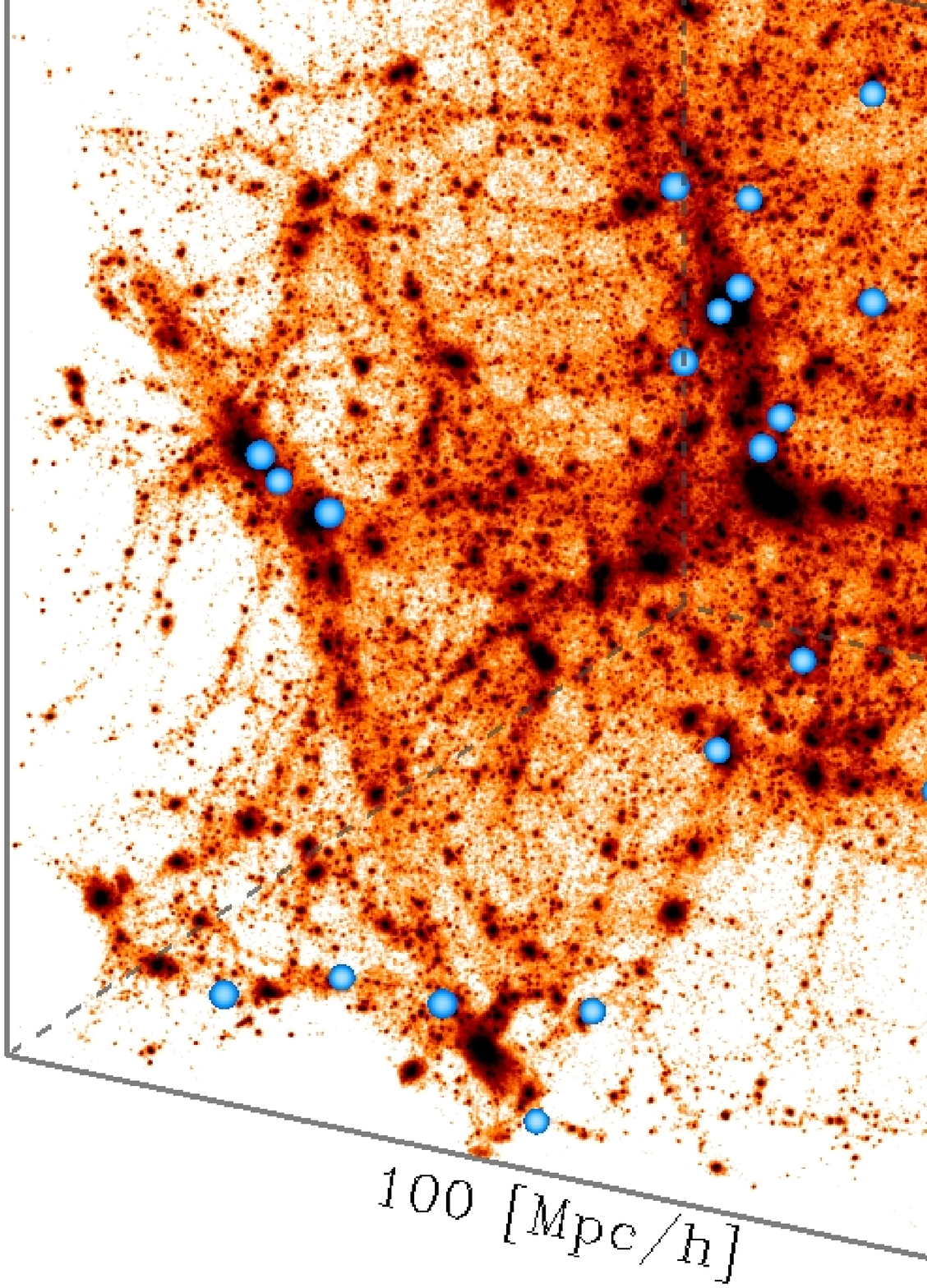}
\caption{The distribution of rotating BHs at $z=0$ in the Millennium simulation. The two panels show the same DM distribution in a cube of comoving length $100~\rm{Mpc~h}^{-1}$ colour coded according to density (black represents the peaks of DM density). Over plotted with blue spheres in the left plot are galaxies with rapidly rotating BHs ($a>0.7$) predicted by the semi-analytic model. Similarly, in the right plot are shown a sample of galaxies with slowly rotating BHs ($a<0.7$) randomly chosen from our data and equal in number to the $a>0.7$ BHs. The size of the spheres is proportional to the spin of the BH that the galaxy hosts.} 
\label{bh_distribution}
\end{figure*}

Below $z=2$ BH mergers start to become an important channel for growing the BH mass, especially when they are between BHs of similar mass, as expected following a major galaxy merger. However, this is a characteristic only of the most massive BHs in our model ($\Mbh>10^9\Msun$). Mergers between those BHs tend to increase the spin of the final remnant to values $a>0.7$ \citep[see also][]{baker_2007, berti_volonteri_2008, fanidakis_2010}. Eventually, at low redshifts BH mergers give rise to a population of rapidly rotating BHs. The appearance of these BHs, as indicated also by the different percentiles below $z=2$, increases the dynamical range of the predicted spins up to values of $0.998$ (the range of spin values in the distribution of BHs in Fig.~\ref{spin_evolution} is smaller because we show only up to the $90^{\rm th}$ percentile of the data). The environmental dependence of BH spins is illustrated in Fig.~\ref{bh_distribution}, where we see that the most rapidly-rotating BHs populate the centres of the most massive DM haloes. In contrast, slowly rotating BHs are found in low mass halo environments. This is a consequence of the correlation between BH mass and spin in our model: rapidly rotating BHs have masses $\gtrsim10^{9}\Msun$. These BHs are hosted by massive elliptical galaxies in our model which inhabit the centres of the most massive haloes.   

In contrast, the median spin of actively growing BHs in our model does not reveal the presence of rapidly spinning BHs. This is demonstrated in the lower panel of Fig.~\ref{spin_evolution} where we have selected the actively growing BHs in our sample ($\dot{m}>0.01$) and plot the median of their spin distribution weighted by disc luminosity. As illustrated in the plot, the predictions are consistent with low spins at all redshifts even at $z\sim0$. This is because in this sample we have selected only the actively growing BHs; we find that almost exclusively these are $\lesssim10^9\Msun$ BH and thus have low spins. Hence, these BHs dominate the sample and determine the over all trend of the median.

\begin{figure}
\center
\includegraphics[scale=0.43]{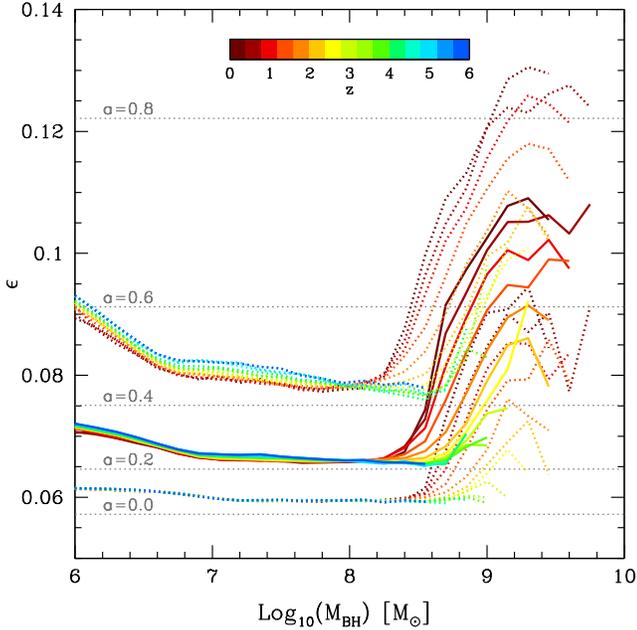}
\caption{The median of the accretion efficiency, $\epsilon$, as a function of BH mass at different redshifts (solid lines) for BHs that accrete in the thin disc regime. Also shown are the $10-90$ percentiles of the $\epsilon-\Mbh$ distribution. We also indicate the values of $\epsilon$ that correspond to spin values of $0, 0.2, 0.4, 0.6$ and $0.8$ (dotted grey lines). The model predicts that low mass BHs have well defined low accretion efficiencies ($\epsilon\simeq0.06-0.08$) at all redshifts. In contrast, in the low redshift universe, a population of very massive BHs with high radiative accretion efficiencies ($\epsilon\gtrsim0.08$) emerges.} 
\label{acc_eff}
\end{figure}

The $a-\Mbh$ correlation can be further elucidated through the dependence  of the accretion efficiency on the mass of actively growing BHs. This is illustrated in Fig.~\ref{acc_eff} where we show the median of the $\epsilon$ distribution for the sample of actively growing BHs (solid lines), again weighted by the disc luminosity. At $z=0$, the median is approximately constant for $\Mbh\lesssim10^{8}$, with a typical value of $\sim0.07$. As implied by the $10-90$ percentiles (dotted lines), the dynamical range of the typical $\epsilon$ values is very small and restricted to the range $0.06-0.08$ ($a=0.1-0.5$). For higher BH masses the efficiency can reach significantly higher values. This is a manifestation of the fact that high mass BHs have high spins and thus high accretion efficiencies when they accrete in the thin-disc regime. Fig.~\ref{acc_eff} also demonstrates that the dependence  of the efficiency on the BH mass does not change with redshift. Hence, at all redshifts BHs have very well determined accretion efficiencies. It is therefore, only the accretion rate that regulates the luminosity output from an accreting BH.

Note that, BHs with higher efficiencies ($\epsilon\gtrsim0.15$) are still found in our BH populations. However, they are usually very massive ($\Mbh>5\times10^8\Msun$) and do not undergo significant quasar activity. These BHs can support the formation of very strong jets in the presence of an ADAF and establish the host galaxy as a radio-loud AGN \citep[high spins and BH mass are essential for high jet luminosities;][]{fanidakis_2010}. In this case, a plot similar to the one in the top panel of Fig.\ref{spin_evolution}, weighted by jet instead of disc luminosity, will unveil a significant population of AGN with rapidly rotating BHs. Hence, BH spins (and efficiencies) can display different distributions depending on the AGN population we are probing. 

\subsection{The distribution of the $\ledd$ parameter}\label{sec:The Ledd parameter}

Having explored the evolution of the BH mass and spin, and calculated the accretion efficiencies for the BHs accreting in the thin-disc regime, we now investigate the disc luminosities of the accreting BHs predicted by the model. We calculate the bolometric disc luminosity in the ADAF and thin-disc regime using the formulation described in Section~\ref{sec:calculation of the disc luminosity}. It is useful to scale $\Lbol$, in units of $\Ledd$ in order to remove the dependence of the luminosity on the BH mass. For this purpose we introduce the Eddington parameter, $\ledd$, defined as,
\begin{equation}
\ledd=\Lbol/\Ledd=\left\{
\begin{tabular}{cl} $\gamma\dot{m}^2,$ & if $\dot{m}<$0.01  \\ 
$\dot{m},$ & if $\dot{m}\geqslant$0.01  \\ 
$\eta[1+\ln(\dot{m}/\eta)],$\negthinspace  &\negthinspace if $\Lbol\geqslant\eta\Ledd$ 
\end{tabular}\right. 
\end{equation}
where $\gamma=0.4\beta/\alpha^2$. 

\begin{figure}
\center
\includegraphics[scale=0.555]{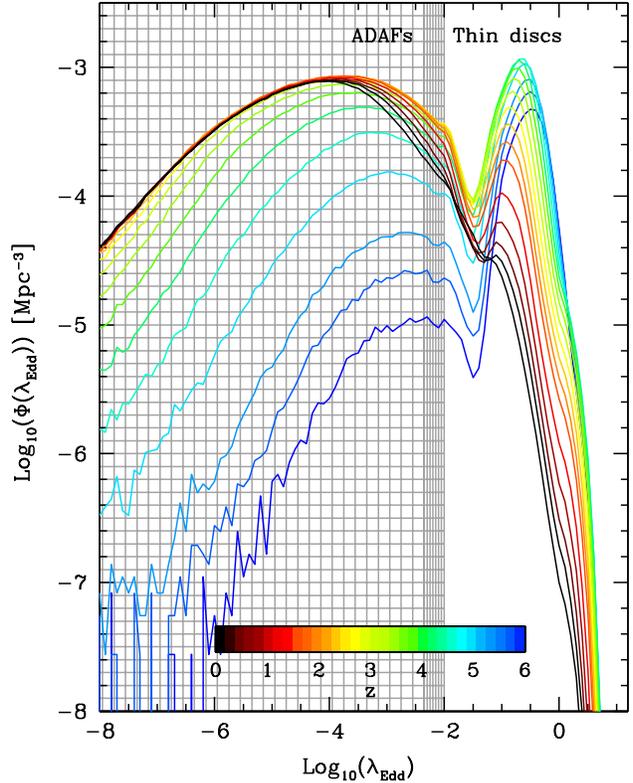}
\caption{The distribution of $\ledd=\Lbol/\Ledd$ for different redshifts as indicated by the colour bar. The plane is divided into the radiative-inefficient regime of ADAFs (shaded region) and radiative-efficient regime of thin discs. The denser shading denotes the region of the plane where the discontinues transition from the ADAF to the thin-disc regime takes place.} 
\label{ledd}
\end{figure}

In Fig.~\ref{ledd} we plot the distribution function of $\ledd$ at different redshifts for all accreting sources with $\Mbh>10^{6}\Msun$. The plane of the plot is divided into two distinct regions: the thin-disc regime and the ADAF regime (note the discontinuity due to the different dependence of $\ledd$ on $\dot{m}$ in the ADAF and thin disc regime). This distinction is essential since it will help us to readily unravel the space density and evolution of luminous and under-luminous AGN in our model. 

The first important property that is unambiguously depicted by Fig.~\ref{ledd} is the bimodal nature of the $\ledd$ distribution. The bimodality is exhibited nearly at all redshifts. The first peak falls in the ADAF regime and it shifts from $\lgledd\simeq-2.5$ at $z\sim6$ to $\lgledd\simeq-4$ at $z\sim0$. The objects that contribute to that mode are all the AGN accreting during the hot-halo mode and those AGN accreting during the starburst mode with $\log\dot{m}<-2$. The second peak is located in the thin-disc regime and in this mode we find the most luminous objects in our model. Their peak space density shifts from $\lgledd\simeq-0.5$ at $z\sim6$ to $\lgledd\simeq-1$ at $z\sim0$, where it nearly disappears. The objects populating the second mode are AGN accreting exclusively in the starburst mode. Since the thin-disc regime is radiatively efficient these objects are expected to dominate the LF of AGN in all bands (except the radio and perhaps the hard X-rays).   
   
The relative space density of the objects in each mode changes dramatically with redshift. At high redshifts, AGN in the thin-disc regime are much more numerous than those in the ADAF regime. The BHs in these AGN have low masses, accrete near the Eddington limit and double their mass several times within a few Gyrs. In contrast, BHs in the ADAF regime are more than an order of magnitude less numerous. Their number density is dominated by BHs that have reached by that time a high mass ($10^8-10^9\Msun$) and their host galaxy undergoes a disc instability or minor merger event that provides gas with a low accretion rate. The first BHs accreting during the hot-halo mode also contribute to the objects populating the ADAF regime. 

This picture changes in the low redshift universe. As redshift decreases more haloes enter the quasi-hydrostatic cooling regime and thus more BHs start to accrete in the ADAF regime (see also the evolution of the hot-halo accretion channel in Fig.~\ref{baryons_SMBHs}). In this mode we also find AGN powered by gas accretion during gas-poor disc instabilities and galaxy mergers. Eventually, at $z=0$ BHs accreting via an ADAF are much more common than those accreting via a thin disc.   

\subsection{The $\Lbol-\Mbh$ correlation}\label{sec:The Lbol-Mbh correlation}
\begin{figure*}
\center
\includegraphics[scale=0.73]{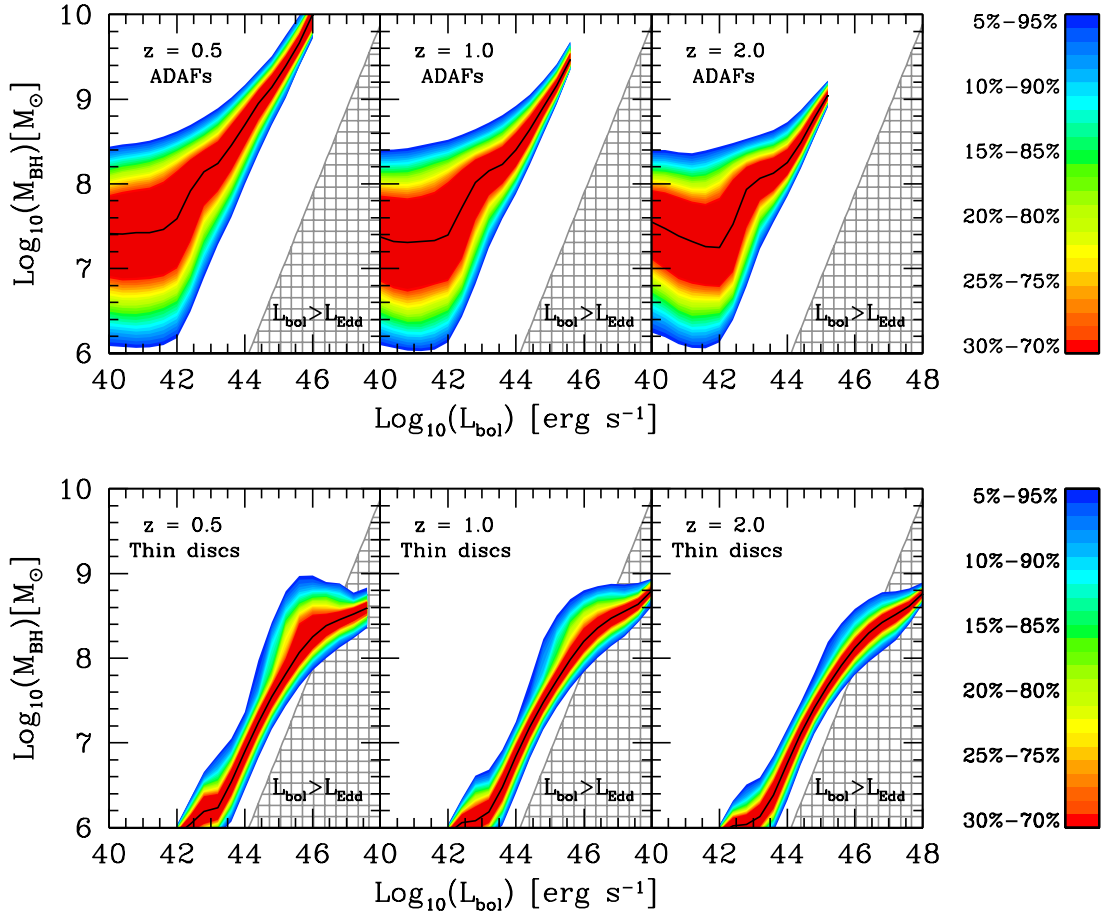}
\caption{The median of the $\Mbh-\Lbol$ distribution at $z=0.5,1$ and $2$ (black solid lines) for BHs accreting via a thin disc (lower panels) and ADAF (upper panels). The different percentiles of the distribution are colour-coded according to the bar on the right. The shaded regions represent the regime where the accretion becomes super-Eddington.}
\label{mbh_lbol}
\end{figure*}

When examining the distribution function of $\ledd$ in Fig.~\ref{ledd} we find a small population of super-Eddington AGN present at all redshifts. The number density of these AGN drops sharply with increasing $\ledd$ due to the logarithmic dependence on the accretion rate. Whether these objects represent the most luminous AGN at some given redshift is not obvious since $\ledd$ is independent of the BH mass. To unravel the relation between the BH mass and accreted luminosity we plot in Fig.~\ref{mbh_lbol} the median of the $\Lbol-\Mbh$ distribution and its associated percentiles at $z=0.5, 1$ and $2$. We show predictions both for objects accreting in the thin-disc (lower panels) and ADAF (upper panels) regime. To guide the reader we also distinguish with different shading the region where $\Lbol\geqslant \Ledd(\Mbh)$. 

When we consider BHs accreting in the ADAF regime we find that the $\Lbol-\Mbh$ correlation is characterized by a floor at luminosities $\lesssim10^{42}\ergsec$. In this regime, BHs have $\ledd\simeq10^{-4}$ and therefore they comprise the majority of accreting BHs in our sample (remember the location of the peak of the $\ledd$ distribution function in Fig.~\ref{ledd}). For example, at $z=0$ we find that $\sim93\%$ of the BHs produce luminosities fainter than $10^{42}\ergsec$. For brighter luminosities, we find a strong correlation between $\Mbh$ and $\Lbol$ which indicates that the most massive BHs must have higher accretion rates compared to the lower mass ones. Indeed, when identifying the accretion properties of the most massive BHs we find that the $\gtrsim10^{8}\Msun$ BHs have $\ledd\gtrsim10^{-3}$. Thus, these BHs are able to produce a significant luminosity even though they accrete in the ADAF regime. In fact, in our sample we find $\sim10^{10}\Msun$ BHs accreting at $\dot{m}\simeq0.01$ which produce luminosities as high as $10^{46}\ergsec$. However, these BHs are very rare: at $z=0.5$, where the space density of these BHs peaks, we find only a handful of them (space densities lower than $10^{-8}~\rm{Mpc}^{-3}$). For higher redshifts, their space density declines and as a consequence the maximum disc luminosity produced in the ADAF regime is also reduced.
 
In the thin-disc regime, we find that accreting BHs typically produce luminosities greater than $10^{42}\ergsec$. The $\log\Lbol-\log\Mbh$ correlation in this regime increases monotonically until the slope becomes significantly shallower near the highest luminosities achieved at a given redshift. The nature of the break in the slope is determined by the AGN feedback prescription in our model. When massive haloes reach quasi-hydrostatic equilibrium they become subject to AGN feedback that suppresses the cooling flows; some of the mass which would have been involved in the cooling flow is instead accreted onto the BH. In these haloes we find the most massive BHs in our model ($\gtrsim10^{9}\Msun$, see Fig~\ref{scal_relations}b). Therefore, these BHs are expected to live in gas poor environments and when they accrete gas they usually do so via an ADAF disc. In this regime, the suppression of cooling flows forces the $\Lbol-\Mbh$ correlation to evolve only along the $\Lbol$ axis since accretion via a thin disc onto $\gtrsim10^{9}\Msun$ BHs becomes very rare. 

The correlation between $\Lbol$ and $\Mbh$ found at $z=0.5$ remains approximately the same at higher redshifts. Only the ranges of BH mass corresponding to the bulk of the AGN shift modestly to lower mass. This is due to the fact that in a hierarchical universe accretion shifts to lower BH masses at higher redshifts (Section~\ref{sec:BH mass evolution}). In addition, the break in the slope at high luminosities becomes less prominent since fewer haloes are in quasi-hydrostatic equilibrium at high redshifts.

The most luminous AGN ($\gtrsim10^{46}\ergsec$) are exclusively powered by super-Eddington accretion onto $\sim10^{8}-10^{9}\Msun$ BHs. This implies that the most luminous quasars are expected to be found in $\sim10^{12}-10^{13}\Msun$ DM haloes and not in the most massive ones (remember the $M_{\rm{Halo}}-\Mbh$ in Fig.~\ref{scal_relations}b). This is in good agreement with the typical DM halo mass quasars are inferred to inhabit as suggested by several observational clustering analyses \citep{da_angela_2008, shen_2009, ross_2009}. In contrast, accretion characterized by lower luminosities spans the whole range of BH masses ($10^{6}-10^{10}\Msun$). 

\section{The evolution of the AGN luminosity functions}\label{sec:The evolution of LFs}

\subsection{Bolometric corrections and obscuration}              

The LF is calculated in redshift ranges which are determined by the observations we are comparing with. The contribution of each AGN to the LF in a range $z=z_1-z_2$ is weighted by a factor, 
\begin{equation}
w_{\rm BH}=t_{\mathrm{active}}/\Delta t_{z_1,z_2},
\label{duty_cycle}
\end{equation}
where $\Delta t_{z_1,z_2}$ is the time interval delineated by the redshift range $z_1-z_2$ and $t_{\mathrm{active}}$ is the time during which the AGN is ``on" in the $\Delta t_{z_1,z_2}$ interval (in principle $t_{\mathrm{active}}= t_{\rm{acc}}$ only if the whole period of accretion falls within $z_1-z_2$). The bands for which we present predictions are the $B$-band ($4400$\AA), soft X-ray (SX: $0.5-2~\mathrm{keV}$) and hard X-ray (HX: $2-10~\mathrm{keV}$). The bolometric corrections considered here are approximated by the following $3^{\rm{rd}}$ degree polynomial relations \citep{marconi_2004},
\begin{align}
&\log(\Lhx/\Lbol)=-1.65-0.22\mathcal{L}-0.012\mathcal{L}^2+0.0015\mathcal{L}^3\nonumber\\
&\log(\Lsx/\Lbol)=-1.54-0.24\mathcal{L}-0.012\mathcal{L}^2+0.0015\mathcal{L}^3\nonumber\\
&\log(\nu_{\rm B}L_{\nu_{\rm B}}/\Lbol)=-0.80+0.067\mathcal{L}-0.017\mathcal{L}^2+0.0023\mathcal{L}^3
\label{bol_corrections}
\end{align}
where $\mathcal{L}=\log(\Lbol/\Lsun)-12$. \citeauthor{marconi_2004} derive these correction based on a spectral template where the spectrum at $\lambda>1\mu m$ is truncated in order to remove the IR bump. In this way, the spectrum corresponds to the intrinsic luminosity of the AGN (optical, UV and X-ray emission from the disc and hot corona). We apply Eqs.~\ref{bol_corrections} to both thin discs and ADAFs, though we note that there is evidence for a change in these corrections with $\ledd$ \citep{vasudevan_fabian_2007}.

There is evidence that the fraction of obscured AGN decreases with increasing X-ray luminosity \citep{ueda_2003, steffen_2003, hasinger_2004, lafranca_2005} a trend found also by \citet{simpson_2005} in a sample of broad and narrow-line AGN from the SDSS. The question of whether the fraction of obscured AGN depends also on redshift is more uncertain. If gas or dust in galaxies provide the obscuring medium then its abundance should evolve in a fashion similar to the SFR history. This will result in a fraction of obscured AGN that is redshift dependent. In addition, a strongly evolving population of obscured AGN is required by AGN population synthesis models to reproduce the properties of the X-ray background \citep{comastri_1995, gilli_1999, ballantyne_2006a, ballantyne_2006b, gilli_2007}. \citet{ueda_2003} and \citet{steffen_2003} suggest that such a trend is not clear in AGN samples of deep X-ray surveys. However, the analysis by \citet{treister_urry_2006} on AGN samples of higher optical spectroscopic completeness indicates that the relative fraction of obscured AGN does increase with redshift. 
 
More recently, \citet{hasinger_2008} showed, based on a sample of X-ray selected AGN from ten independent samples with high redshift completeness, that the fraction of obscured AGN increases strongly with decreasing luminosity and increasing redshift. According to \citeauthor{hasinger_2008}, the dependence  of the obscured fraction of AGN, $f_{\rm{obsc}}$, on $\Lhx$, can be approximated by a relation of the form, 
\begin{equation} 
\fobsc=-0.281\log(L_{\mathrm{HX}})+A(z),
\label{f_obsc}
\end{equation}
where $A(z)=0.308(1+z)^{0.48}$ and $\Lhx=10^{42}-10^{46}\ergsec$. A broken power law can also be used to describe $A(z)$ which provides a fit with modestly better statistical significance. The best fit gives a power law of the form $A(z)\propto(1+z)^{0.62}$ that saturates at $z=2.06$ and remains constant thereafter. 

Hence, in order to determine the correct population of AGN in a luminosity bin of a given band we need to take into account the effects of obscuration. To do so, we utilise the dependence  of the obscured fraction on the $\Lhx$ luminosity found in \citet{hasinger_2008} as follows. We calculate the fraction of visible AGN, $\fvis=1-\fobsc$, at a given luminosity bin in the $2-10~\mathrm{keV}$ band using Eq.~\ref{f_obsc} and then we associate the value of $\fvis$ with the corresponding B-band or soft X-ray luminosity bin using Eqs.~\ref{bol_corrections}. The LF can then be expressed as,
\begin{equation}
\frac{d\Phi}{d\log(L_X)}\bigg\vert_{\rm{vis}}=\fvis(\Lhx,z)\frac{d\Phi}{d \log(L_X)}.
\end{equation}
In our analysis, we choose the single instead of the broken power law form of $A(z)$ to allow for a comparison with previous work \citep[e.g.,][]{lafranca_2005}. In this way we provide a simple, yet well constrained, prescription for the effects of obscuration in  the AGN of our model. 

\subsection{The optical LF}\label{sec:The optical LF}

\begin{figure*}
\center
\includegraphics[scale=0.7]{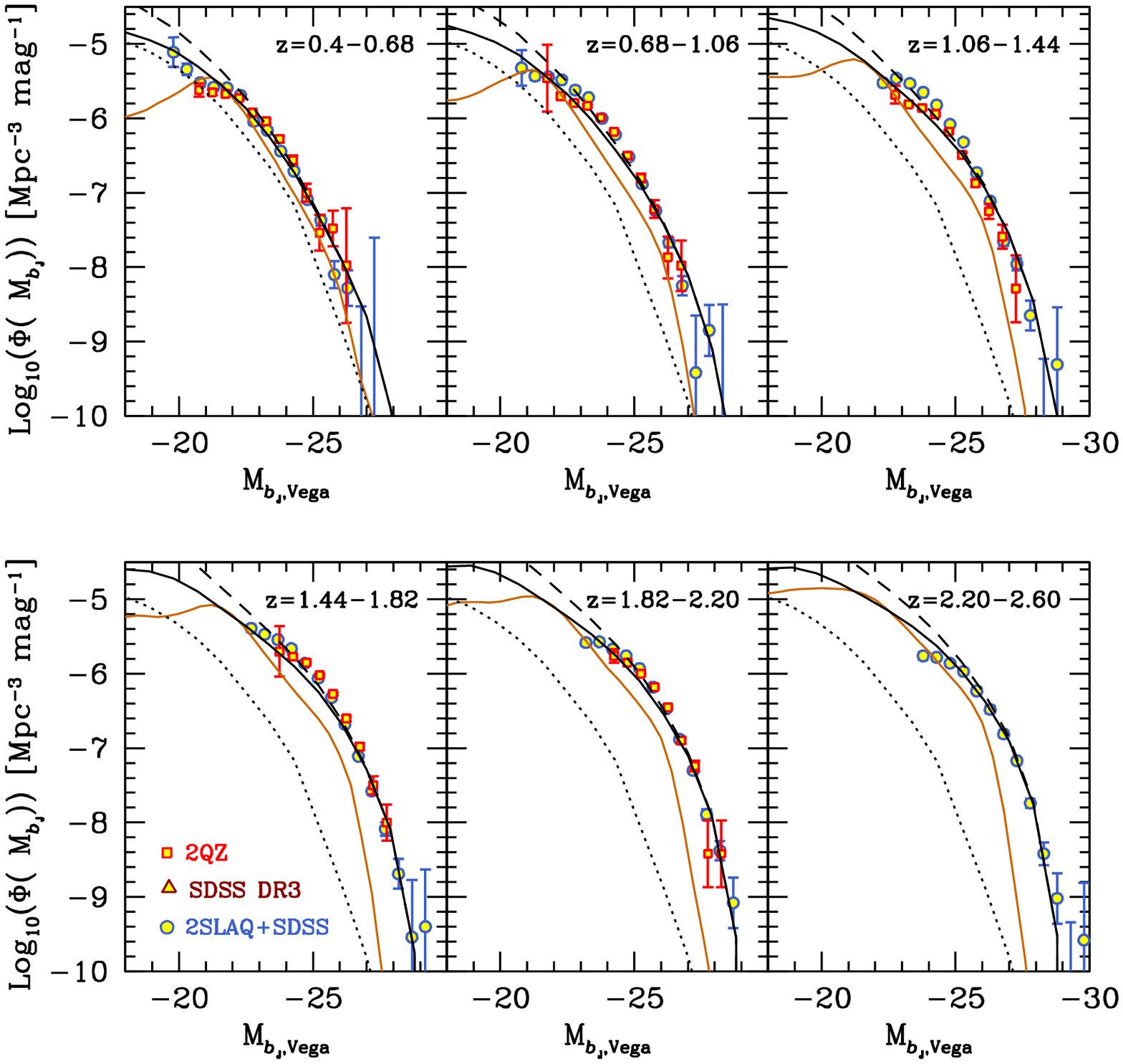}
\includegraphics[scale=0.7]{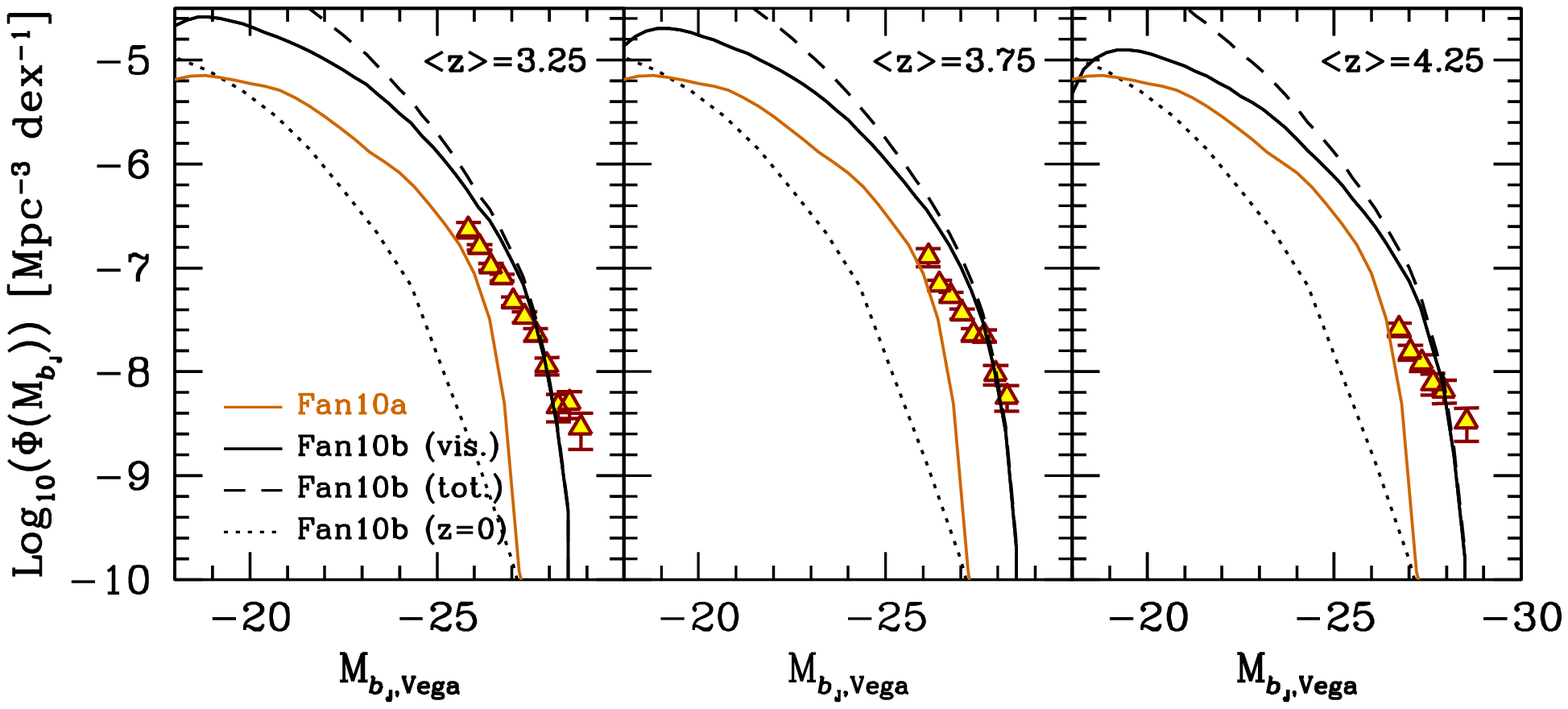}
\caption{The $\bj$-band quasar LF in the redshift range $0.4<z<4.25$. Predictions are shown for the model described in this paper (Fan10b; solid black lines) and the Fan10a (solid orange lines) model. For reference, we show the $z=0$ LF in the Fan10b model (dotted black line). We also show the evolution of the LF in the Fan10b model before applying the obscuration (dashed black lines). The observations represent the LFs estimated from the 2SLAQ+SDSS \citep[blue-filled circles,][]{croom_2009b}, 2QZ \citep[red-filled circles,][]{croom_2004} and SDSS \citep[brown-filled triangles,][]{richards_2006} surveys. } 
\label{qlf_bj}
\end{figure*}

In Fig.~\ref{qlf_bj} we present the $\bj$-band LF of quasars in 9 different redshift bins between $0.4<z<4.25$. Our predictions are shown before and after applying the obscuration effect (dashed and solid black lines respectively).  Absolute magnitudes are first calculated in the B band using the standard expression
\begin{equation}
M_{\mathrm{B}}=-10.44-2.5\log(L_{\mathrm{B}}/10^{40}\mathrm{erg~s^{-1}}),
\end{equation}
for magnitudes in the Vega system and then converted to the $\bj$ band using the correction $\Mbj=M_{\rm B}-0.072$ \citep{croom_2009b}. Predictions are shown for the Fan10b and for comparison, the predictions from the modified version of the \citet{bower_2006} model (solid orange lines) presented in \citet[][Fan10a]{fanidakis_2010}. The \citeauthor{fanidakis_2010} model was constrained to match the observed LF of quasars at $z\sim0.5$. In brief, the model assumes a constant obscuration fraction of $\fvis=0.6$ and a $\bj$-band bolometric correction of $0.2$. Finally, our predictions for the $z=0$ LF, after applying the obscuration effect, are shown in every redshift bin (black dotted lines) to help us assess the evolution with redshift. 

Our predictions are compared to the 2SLAQ+SDSS QSO LFs \citep{croom_2009b}. QSO magnitudes in the 2SLAQ sample were obtained in the $g$-band and therefore need to be converted to the $\bj$ band considered here. We use $\Mbj=M_{g}+0.455$ \citep{croom_2009b} where the K-corrections for the $g$ band have been normalized to $z=2$. In addition, we plot the 2QZ LFs by \citet{croom_2004}. The observed LFs agree very well with each other particularly at bright magnitudes ($\Mbj<-24$). The modest disagreement seen at the faintest magnitudes is due to the different K-correction applied to the 2QZ QSO sample by \citet{croom_2004}. For higher redshifts, we include the LF derived from the SDSS Data Release 3 sample \citep[DR3,][]{richards_2006}. The SDSS DR3 LF is obtained in the $i$ band, therefore we need to convert it to the $\bj$ band. We use $\Mbj=M_{i}+0.71$, assuming a spectral index of $a_{\nu}=-0.5$ \citep{croom_2009b}.

The AGN model contains one free adjustable parameter that must be constrained by observational data. This is the $\fq$ parameter in Eq.~\ref{acc_timescale} which sets the accretion timescale. We choose the $\bj$ band for constraining the value of $\fq$ because it is the most sensitive to the modelling of AGN.  Other free parameters are constrained by the galaxy formation model and are adjusted to reproduce the observed galaxy LFs \citep[][see also \citealt{bower_2010} for further discussion of our parameter fitting philosophy]{bower_2006, lagos_2010}. Since $\fq$ is essentially the only free parameter in our model, the predictions presented in these Sections are genuine predictions of the underlying galaxy formation model. By fixing $\fq$ to $10$ we obtain an excellent overall match to the observed QSO LFs in the $0.4<z<2.6$ redshift interval. For higher redshifts, our model provides a good match, however, it predicts a much steeper slope for the bright end compared to the SDSS DR3 LF. 

\begin{table}
\begin{center}
\caption{Typical values of $\fvis$ in three magnitude bins and its evolution with redshift.}
\label{obscuration}
\begin{tabular}{@{}lccccc}
\hline
\hline
 & $\Mbj=-20$ & $\Mbj=-24$ & $\Mbj=-28$ \\
\hline
 $z=0.4$ 	& 53.3\% 	& 84.3\% 	& 100\% \\
 $z=0.68$	& 49\% 	& 81\% 	& 100\% \\
 $z=1.06$	& 44.9\%	& 76.9\% 	& 100\%\\
 $z=1.44$	& 42.2\% 	& 73.3\% 	& 100\%\\
 $z=1.82$	& 27.8\% 	& 69.9\% 	& 100\%\\
 $z=2.20$	& 24.7\% 	& 66.7\% 	& 98.7\%\\
\hline
\hline
\end{tabular}
\\
\end{center}
\end{table}

A comparison between the predictions for the $z=0.4-4.25$ and $z=0$ LFs in Fig.~\ref{qlf_bj} shows that quasars undergo significant cosmic evolution. For example, quasars with $\Mbj=-25$ increase in space density from $\sim10^{-8}$ to $\sim10^{-6}~\rm{Mpc}^{-3}$ between $z=0$ and $z=2.2-2.6$. This strong evolution in the space density of quasars is due to the fact that disc instabilities and galaxy mergers, the two processes that trigger accretion during the starburst mode, become more frequent at higher redshifts (see Fig.~\ref{sfr}). Yet, the strong evolution of quasars does not affect only their space density. At a fixed space density of $10^{-8}~\rm{Mpc}^{-3}$ the quasar LF brightens from $\Mbj\simeq-25$ at $z=0$ to $\Mbj\simeq-28$ at $z=2.2-2.6$. This is attributed to the fact that the cold gas becomes more abundant with increasing redshift. This is equivalent to a strong increase in the gas reservoir available for feeding the central BHs since more gas is turned into stars when a starburst is triggered. 

The processes that are responsible for driving the formation of stars in starbursts are also responsible for the cosmic evolution of quasars. Qualitatively, the strong link between the formation of stars and quasar activity (and therefore BH growth) can be illustrated by considering the Fan10a model. In Fig.\ref{sfr} we can see that the SFR density in bursts in the Fan10a model shows a milder evolution with redshift compared to the Fan10b model. Hence, the model predicts considerably that less gas is available for accretion which then results in more modest evolution of the quasar LF. As a consequence, the predictions of the model Fan10a provide an overall poor match to the observed LFs. 

The evolution of the SFR density with redshift does not imply a proportional evolution of quasar luminosities on the $\Mbj-\Phi(\Mbj)$ plane. The space density of the brightest quasars increase with redshift. However, the density of objects around the break in the LF increases more quickly, leading to a steepening of the LF as illustrated by Fig.\ref{qlf_bj}. This differential evolution is determined exclusively by the accretion physics in our model. In the $0.01<\m<1$ regime, the disc luminosity scales in proportion to $\m$. However, when the flow becomes substantially super-Eddington, the luminosity instead grows as $\ln(\m)$ and therefore the dynamical range of predicted luminosities decreases dramatically. The logarithmic dependence of luminosity on the accretion rate has a strong impact on the shape of the LF, resulting in a very steep slope at the bright end. This becomes apparent in the highest redshift intervals ($z>1.44$) where the relative number of super-Eddington accreting sources becomes significantly higher than in the lower redshift intervals (because more gas is available for accretion, see Fig.~\ref{ledd}). Since these sources exclusively populate the brightest magnitude bins, the bright end of the LF now becomes significantly steeper.

Another factor that influences the evolution of the quasar LF is obscuration. Low luminosity quasars are heavily obscured according to our obscuration prescription, and thus remain well buried in their host galactic nuclei. This is clear from Table \ref{obscuration} where we summarize the evolution of the $\fvis$ value in the redshift range of interest. In the highest redshift intervals the obscuration becomes more prominent affecting even the brightest sources. However, the intense accretion activity during the starburst mode dominates the evolution of the faint end, and therefore a strong cosmic evolution in the space density of the faintest sources, similar to that of the brightest sources, is still observed. Nonetheless, the obscuration significantly influences the cosmic abundance of the quasar populations dominating the faint end, something that will become more evident in Section~\ref{sec:The evolution of cosmic AGN abundances} where we will explore the cosmic evolution of quasars of given intrinsic luminosity.
 
Finally, we point out that those AGN that are powered by an ADAF contribute significantly to the space density of AGN fainter than $\Mbj\simeq-22$ in the low redshift universe. In fact, in the range $0.4<z<0.68$, we find ADAF sources with $\dot{m}\simeq0.001-0.01$ contributing also to the knee of the LF. These systems represent the rare $\gtrsim10^{9}\Msun$ accreting BHs in Fig.~\ref{mbh_lbol}. Identifying the ADAF systems with optically bright quasars introduces a caveat for the model since quasars typically show high excitation spectra which indicate the presence of a bright, UV thin disc \citep[see e.g.,][]{marchesini_2004}. Nonetheless, these systems accrete where the transition to a thin disc takes place. Observations of stellar mass BH binary systems show that this transition is complex, probably taking on a composite structure with the thin disc replacing the hot flow at progressively smaller radii \citep[see e.g,. the review by][]{done_2007}. Such a configuration could possibly produce high excitation lines while retaining also the ADAF characteristics. For a more comprehensible representation of the objects populating the LF, we refer the reader to Fig.~\ref{blf} in Section~\ref{sec:The bolometric LF} where we decompose the LF into the contribution from ADAF, thin-disc, and also super-Eddington sources.

\subsection{The X-ray LFs}\label{sec:The X-ray LFs}
\begin{figure*}
\center
\includegraphics[scale=0.7]{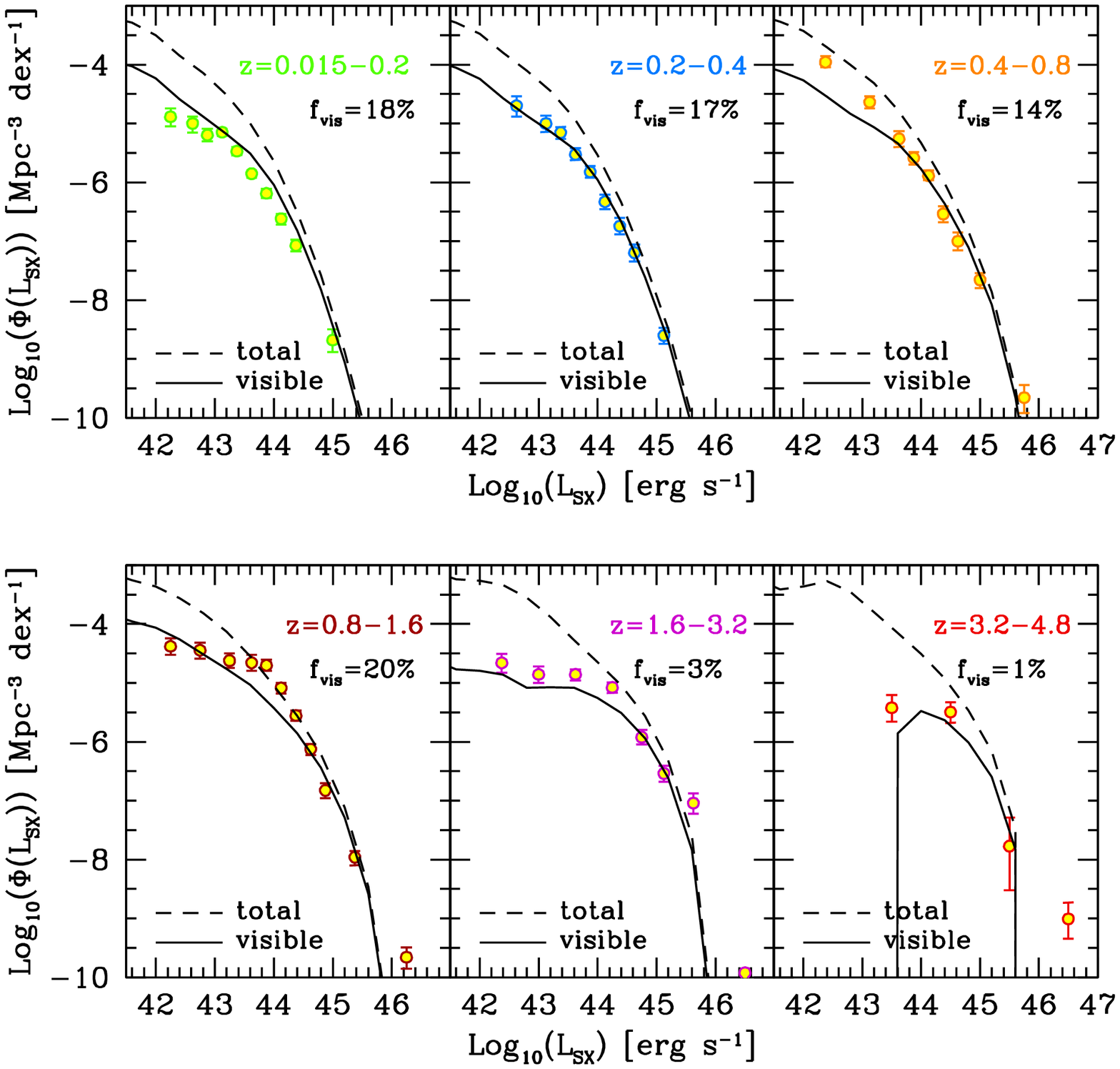}
\caption{The predictions of our model for evolution of the soft X-ray LF (solid black lines). Predictions are shown before (total: dashed lines) and after (visible: solid lines) we apply the effects of obscuration. In addition, we quote the total fraction of AGN with $L_{\rm SX}>10^{42}\ergsec$ that are visible in every luminosity bin. Data are taken from \citet{ueda_2003}.} 
\label{qlf_sxr}
\includegraphics[scale=0.7]{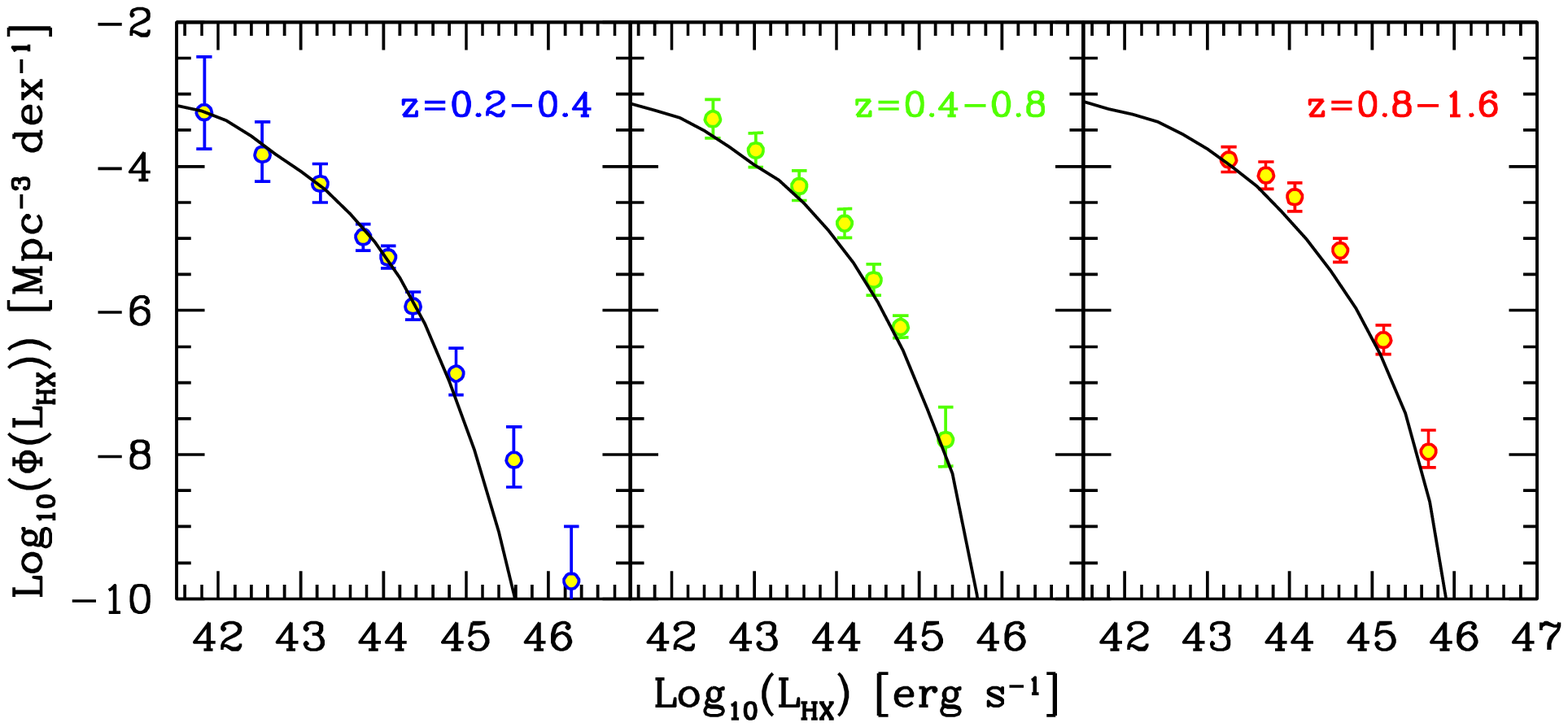}
\caption{The evolution of the hard X-ray LF in our model (solid black lines). Predictions are shown only for $z<1.6$ and no obscuration correction is applied. Observational data are taken from \citet{hasinger_2005}.} 
\label{qlf_hxr}
\end{figure*}  
The good agreement between our model predictions for the optical LF and the observations motivates us to study also the evolution of X-ray AGN as a function of redshift and intrinsic luminosity. X-rays account for a considerable fraction of the bolometric luminosity of the accretion disc and therefore provide an ideal band for studying the cosmic evolution of AGN. The observed continuum spectra of AGN in the X-rays can be represented by a simple power law in the \emph{hard} part of the spectrum ($2-10~\rm{keV}$) with a universal spectral index of $\alpha_X\sim0.7$ which is independent of the luminosity over several decades. In the \emph{soft} part of the spectrum ($0.5-2~\rm{keV}$) the power law continuum is strongly attenuated by absorption which is typically in excess of the Galactic value and corresponds to a medium with hydrogen column densities of $N_{\mathrm{H}}\simeq10^{20}-10^{23}~\mathrm{cm}^{-1}$ \citep{reynolds_1997, george_1998, piconcelli_2004}. 

The cosmic evolution of X-ray AGN has been investigated by employing AGN selected both in the soft \citep{miyaji_2000, hasinger_2005} and hard part of the spectrum \citep{ueda_2003, lafranca_2005, barger_2005, aird_2010}. Mirroring the strong evolution of the optical LF, the LF of X-ray selected AGN also evolves strongly with cosmic time. The LFs in soft and hard X-rays have similar characteristics, evolving at the same rate but differ significantly in normalization. For example, the \citet{hasinger_2005} soft X-ray sample has approximately a 5 times smaller global LF normalization compared to that of the hard X-ray LF estimated by \citet{ueda_2003}. This difference is most likely attributed to the fact that a large fraction of AGN is obscured in the soft X-rays \citep[type-2 AGN are by a factor of 4 more numerous than type-1 AGN,][]{risaliti_1999}.

We present in this section our predictions for the X-ray LFs. We calculate the X-ray emission in the $0.5-2~\mathrm{keV}$ and $2-10~\mathrm{keV}$ bands using Eq.~(\ref{bol_corrections}). Our predictions for the soft and hard X-ray LFs are shown in Figs.~\ref{qlf_sxr} and \ref{qlf_hxr} respectively (solid lines). The predictions are compared to the LFs estimated by \citet{hasinger_2005} for the soft X-rays and \citet{ueda_2003} for the hard X-rays. Note that we do not show estimates from observations that consist only of upper limits (empty bins). The \citeauthor{hasinger_2005} sample comprises unabsorbed AGN and therefore we need to take this into account for the sample in the soft X-rays the effect of obscuration. We do so by using the \citet{hasinger_2008} prescription as explained in the previous section. To quantify the effect of absorption we also plot in the soft X-rays the total population of AGN (dashed lines). 

Overall, our model provides a very good match to the observations in both soft and hard X-rays. We find discrepancies between the observations and the model predictions in the redshift range $z=0.4-0.8$, where the space density of the $<10^{43}\ergsec$ soft X-ray AGN is under predicted. A similar disparity is also seen in the $z=0.8-1.6$ bin for the $10^{43}-10^{45}\ergsec$ AGN. However, it is important to bear in mind that our predictions in the soft X-rays depend strongly on our prescription for the obscuration. Since our predictions for the total AGN population are always well above the observations, it is likely that these discrepancies are attributed to the modelling of the obscuration. We also find that the model predicts at high redshifts a somewhat steeper bright end compared to the observations, although the very brightest point could be due to beaming effects in a small fraction of the more numerous lower luminosity sources \citep{ghisellini_2010}. Nonetheless, the very steep slope in our predictions is again a manifestation of the Eddington limit applied to the super-Eddington sources, an effect that becomes more significant at higher redshifts as already explained in the previous section.

As illustrated by the predictions for the soft X-ray LF in Fig.~\ref{qlf_sxr}, the space density of the faintest soft X-ray AGN in our model is strongly affected by the obscuration. As illustrated by the LFs in the soft X-rays, AGN with $L_{\rm{SX}}\simeq10^{42}-10^{44}\ergsec$ are heavily obscured. In fact, in the redshift interval $z=0.015-0.2$ only $19\%$ of the total AGN population with $L_{\rm{SX}}\geqslant10^{42}\ergsec$ is visible in soft X-rays. Hence, the obscured AGN outnumber the visible AGN by a factor of 4 \citep{risaliti_1999}. This fraction of visible AGN becomes even smaller at higher redshift because the obscuration becomes more prominent with redshift. At $z>3.2$ only a negligible fraction of the AGN are seen in soft X-rays. Based on these results we estimate that approximately $10\%$ of the total number of AGN with $L_{\rm{SX}}\geqslant10^{42}\ergsec$ are visible in the soft X-rays in the $z=0.015-4.8$ universe. We note that these estimations are based on the obscuration model which according to the AGN sample from \citet{hasinger_2008} is evaluated only for luminosities $10^{42}\ergsec<L_{\rm{HX}}<10^{46}\ergsec$. For luminosities lower than $10^{42}\ergsec$ we assume that $\fvis$ remains constant and equal to $\fvis(10^{42}\ergsec)$. It is therefore likely that we underestimate the value of $\fvis$.

\subsection{The bolometric LF}\label{sec:The bolometric LF}
\begin{figure*}
\center
\includegraphics[scale=0.7]{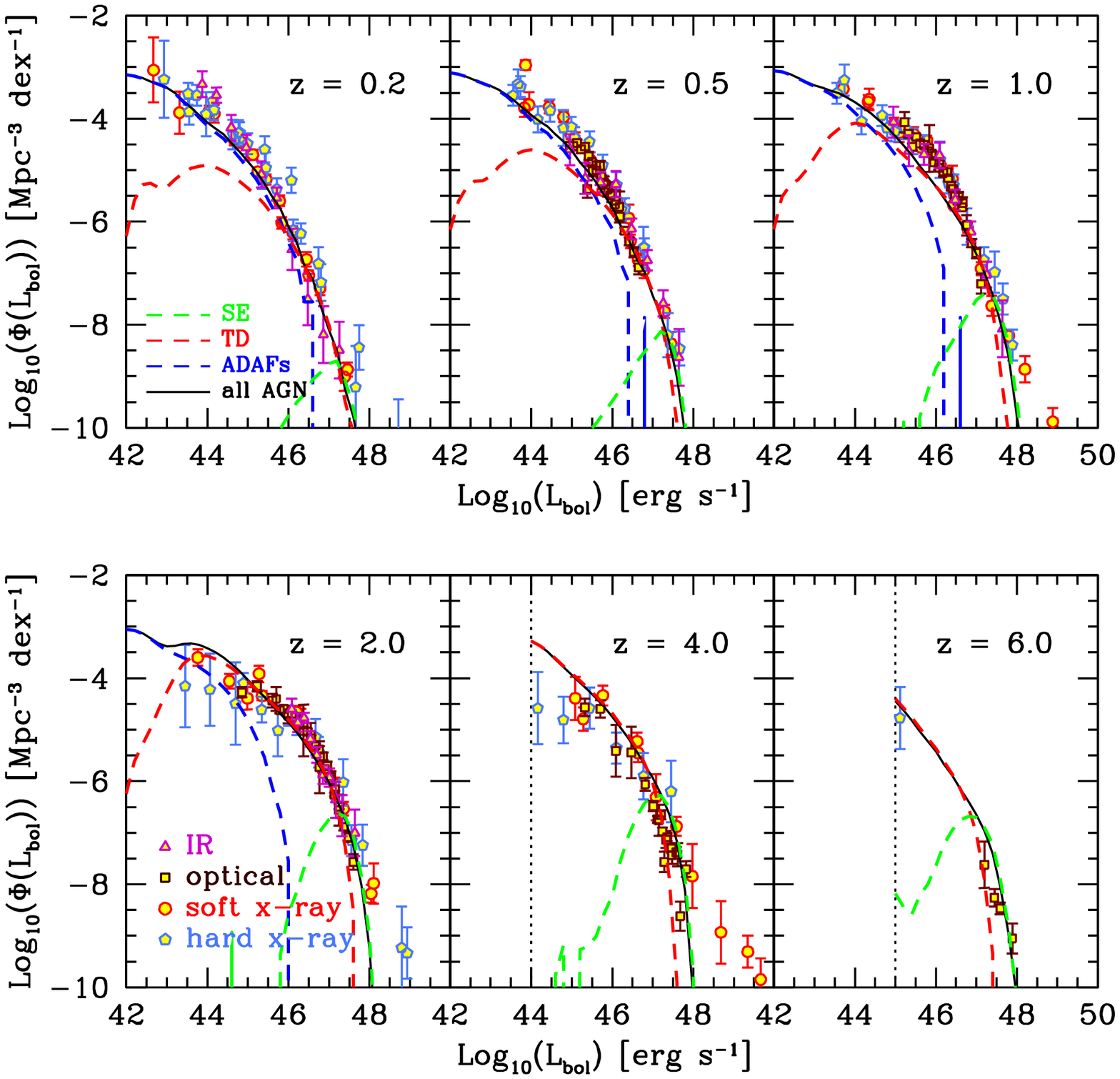}
\caption{The bolometric LF predicted by our model in 6 redshift bins between $z=0.2$ and $z=6$. Also shown is the bolometric LF estimated by \citet{hopkins_2007} using a large sample of infrared (IR) optical, soft and hard X-ray LFs (we refer the reader to \citealt{hopkins_2007} for a description of the observational samples used by the authors). The different bands are depicted by different colours as indicated by the labels. The dashed lines show the contribution to the LF of the ADAF (blue), thin disc (red) and super-Eddington sources (green). The vertical dotted lines indicate the resolution limit of the Millennium simulation.} 
\label{blf}
\end{figure*} 

We have so far studied the LF of AGN and quasars in different wavebands using the bolometric corrections in Eq.~(\ref{bol_corrections}) and compared them to the available observations. We find that our predictions match reasonably well the observations, particularly in the $\bj$ band. Further insight into the evolution of AGN can be gained by calculating the bolometric LF and studying its evolution with redshift. This will provide an overall characterisation of the global AGN population and might reveal additional trends which are perhaps unseen when exploring the LF in the other bands we studied earlier. 

To calculate the bolometric LF we consider all accreting objects (in both the starburst and hot-halo mode) taking into account the regime these objects accrete in (ADAF or thin disc). This time we probe a wider baseline in luminosity and thus we expect to see clearly the contribution of the AGN powered by ADAFs. We calculate the bolometric LF at some redshift $z$ following the technique described in the previous sections. AGN are sampled over a period equal to $20\%$ of the age of the universe at redshift $z$. We need to stress that the resolution of the Millennium simulation has an impact on the gas properties of the $\lesssim10^{44}\ergsec$ objects at high redshifts. Therefore, we compare the N-body results with high-resolution simulations using Monte-Carlo (MC) halo merger trees to test the limit of our predictions. The MC algorithm we use to generate the DM halo merger trees has been presented in \citet{parkinson_2007}. The algorithm is a modification of the Extended Press-Schechter algorithm described in \citet{cole_2000} and accurately reproduces the conditional MFs predicted by the Millennium N-body simulation. The comparison shows that the calculation with N-body trees becomes incomplete for $\Lbol\lesssim10^{44}-10^{45}\ergsec$ AGN at $z\gtrsim4$. Therefore, at high redshifts we show predictions only down to the resolution limit of the Millennium simulation (the limit is indicated by the vertical dotted lines). 

Our predictions for the bolometric LF are shown in Fig.~\ref{blf} (solid black lines) in the redshift range $z=0.2-6$. In addition, we show independently the contribution to the LF from the ADAF, thin-disc, and $\Lbol\geqslant\Ledd$ sources (dashed coloured lines). Our results are compared to the bolometric LF extracted from observations by \citet{hopkins_2007}, using a combination of a large set of observed LFs in the optical, X-ray, near- and mid-infrared wavebands. The authors are able to reproduce the bolometric LF and the individual bands by employing their best-fit estimates of the column density and spectral energy distribution, but also a prescription for the obscuration fraction which is assumed to be a function of the luminosity (though not of redshift).  Their AGN LF also show the downsizing of AGN activity at low redshifts, which is expressed through the steepening of the faint-end slope below $z=1$. The data used by \citeauthor{hopkins_2007} cover redshift wide redshift ranges and sometimes the same data sets appear in more than one bin (especially the soft X-ray data at $z>0.8$). Therefore, comparison between the data and our predictions should be treated with caution because the observations over such wide redshift bins could hide evolutionary trends. 

Nevertheless, our predictions for the bolometric LF match very well the observational estimates across a wide range of redshifts. In the low-redshift universe ($z\lesssim1$), the model reproduces the faint and bright ends of the observed LF remarkably well. At the faint end, the model predicts a significant contribution from AGN powered by ADAFs. These are predominately massive BHs accreting at low accretion rates during the hot-halo mode and they account for the increasing abundance with decreasing redshift of the faint AGN (see also Fig.~\ref{ledd}). By contrast, the bright end is always populated by AGN radiating near or greater than the Eddington limit. These are the AGN that harbour the rapidly growing $10^{7}-10^{8}\Msun$ BHs in the low-redshift universe. 

At high redshifts ($z\gtrsim1$), the contribution from the ADAF sources becomes less important. This is because not only fewer haloes reach quasi-hydrostatic equilibrium, but also most of the accretion during the starburst mode takes place in the thin-disc regime. The LF at these redshifts is dominated by AGN whose BHs accrete during the starburst model. The BHs in these AGN grow very fast as they usually double their mass several times during a single accretion episode. This gives rise to a strongly evolving population of super-Eddington sources that dominate the bright end of the LF. 

\begin{figure}
\center
\includegraphics[scale=0.43]{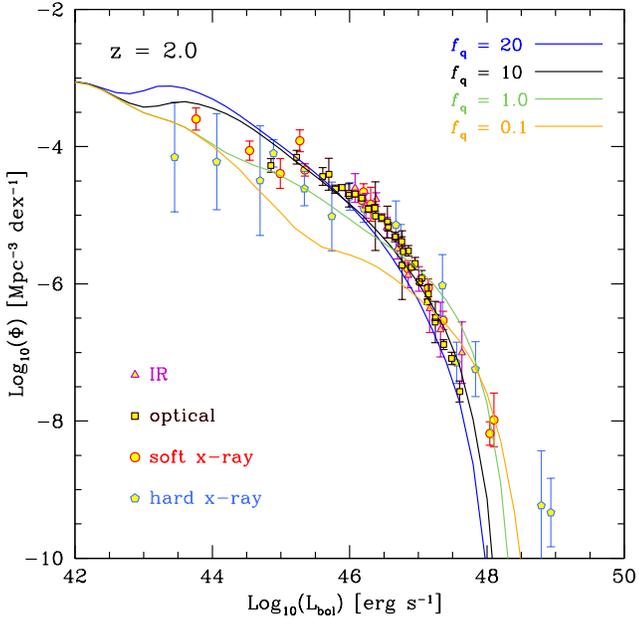}
\caption{Our predictions for the bolometric LF at $z=2$ considering four different values of the accretion timescale parameter,$\fq$, as indicated in the legend. Data are taken from \citet{hopkins_2007}.} 
\label{blf_tacc}
\end{figure}      

The bright end of the $z\geqslant1$ bolometric LF has a very steep slope compared to the observational estimate, indicating a deficiency of very bright quasars ($\Lbol>10^{48}\ergsec$). The inability of the model to predict the luminosities of the most luminous quasars is due to the fact that the accretion luminosity is constrained by the Eddington limit. As discussed before, applying the Eddington limit to the accreting sources in our model results in a very steep decline in the space density of the super-Eddington sources, an effect clearly seen earlier in the optical LF. The reader needs to keep in mind though that, as far as the observational estimates are concerned, many of the brightest luminosity bins do not include any objects and therefore, they constitute only upper limits.

To achieve higher quasar luminosities we need to increase the accretion rates in the most massive BHs to values much higher than the Eddington rate\footnote{For comparison, a BH accreting at $\m\sim100$ is only $\sim5$ more luminous that a BH accreting at $\m=1$.}. This is necessary because even if we assume that all our BHs accrete at the Eddington limit the maximum luminosity we can produce is $\sim10^{48}\ergsec$ (when we consider, for example, a $10^{10}\Msun$ BH) which is still lower than the highest luminosities observed. Note though that, AGN feedback prevents BHs more massive than $10^{9}\Msun$ from accreting at or near supper-Eddington accretion rates in our model.

To increase the accretion rate of a BH we can either increase the amount of gas that is fed into it or decrease the accretion timescale. The former solution corresponds to increasing the value of $f_{\mathrm{BH}}$ (the parameter that determines the fraction of gas available for accretion) which has already been tuned to provide a good match to the local BH density and MF. The latter solution gives us the freedom to adjust the accretion rates without changing the BH mass properties. By decreasing the value of $\fq$ in Eq.~\ref{acc_timescale} we can obtain higher $\m$ values and therefore boost the produced bolometric luminosities to $\Lbol>10^{48}\ergsec$. This is illustrated in Fig.~\ref{blf_tacc}, where we show the bolometric LF at $z=2$ assuming $\fq=20,10,1$ and $0.5$. However, changing the value of $\fq$ has a strong effect on the faint end of the LF since it decreases the space density of the low accretion rate objects (the space density of the $\lesssim10^{43}\ergsec$ AGN remains unchanged since that part of the LF is dominated by AGN in the hot-halo mode in which the accretion timescale is derived directly from the cooling timescale of the gas). In addition, it reduces significantly the duty cycle of actively growing BHs by limiting the typical accretion timescale to $\sim10^5-10^6$~yr if $\fq\leqslant1$. Hence, having a constant value of $\fq$ for all the accreting BHs limits our ability to account for the entire $\Lbol$ baseline. Perhaps this suggests that a more sophisticated treatment of the accretion timescale is therefore needed. However, this is beyond the scope of the present analysis. 

\subsection{The evolution of cosmic AGN abundances: cosmic downsizing?}\label{sec:The evolution of cosmic AGN abundances}
\begin{figure*}
\center
\includegraphics[scale=0.57]{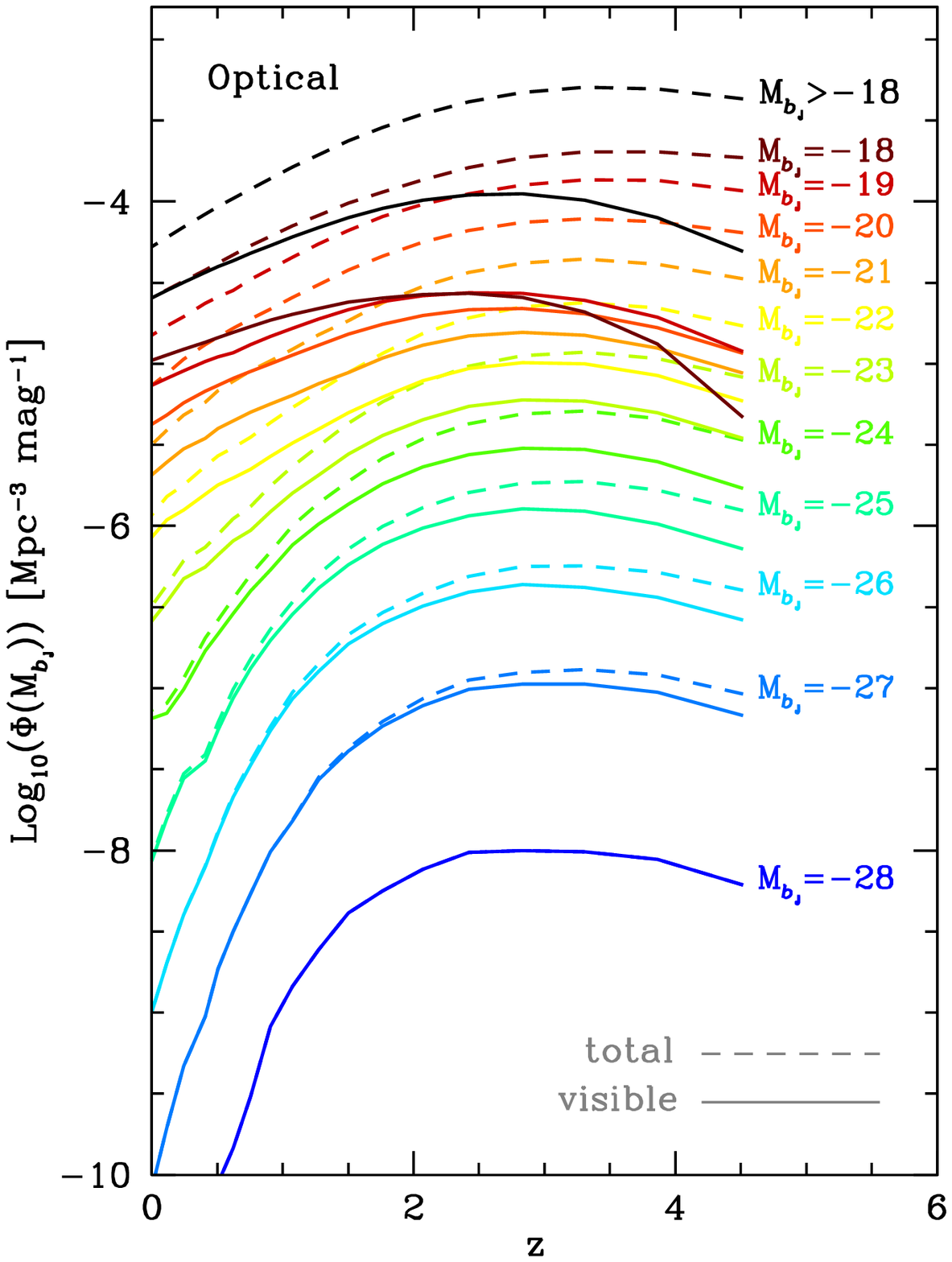}
\includegraphics[scale=0.57]{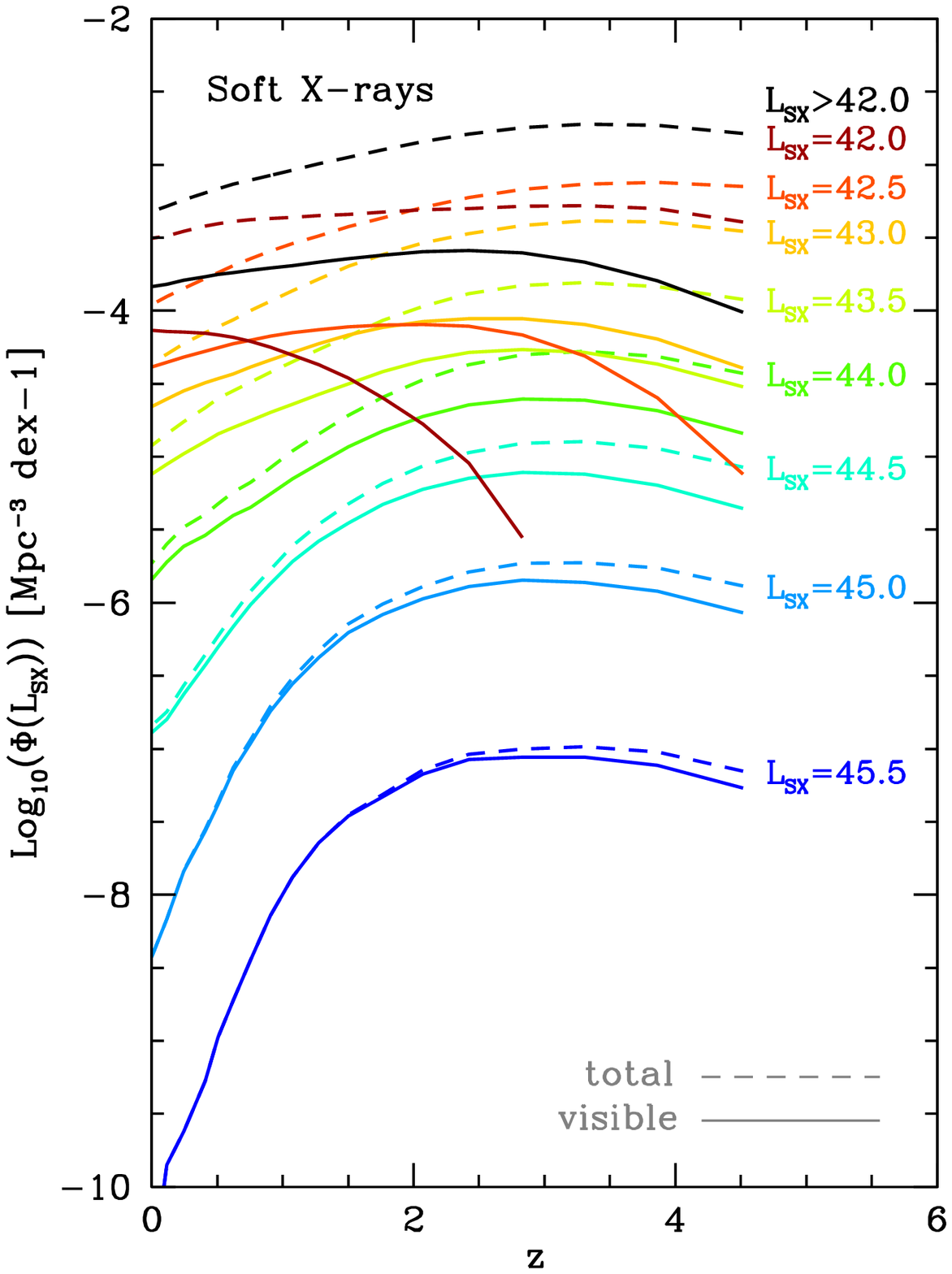}
\caption{The cosmic evolution of different magnitude and luminosity classes of AGN selected in the optical $\Mbj$-band (left panel) and soft X-rays (right panel) respectively. In addition, we show the cosmic evolution of the sum of all AGN with $\Mbj\leqslant-18$ in the optical and $\log L_{\rm{SX}}\geqslant10^{42}\ergsec$ in the soft X-rays (black lines). In all cases we show with dashed and solid lines the evolution of the space density of AGN before (total) and after (visible) applying the effects of obscuration.} 
\label{downsizing}
\end{figure*}        

Our predictions for the optical, X-ray and bolometric LFs, suggest that AGN undergo important cosmic evolution. Their space density was significantly higher at earlier epochs, an evolutionary trend which suggests that AGN activity in the past was much more intense. Interestingly, our predictions indicate that the effects of the different physical processes considered here (the two accretion modes and the obscuration prescription) must have an important, and indeed variable, effect on the evolution of faint and bright AGN. Therefore, to gain more insight into the cosmic evolution of AGN we study how the abundance of different luminosity populations of AGN evolves with redshift. This is shown in Fig.~\ref{downsizing}, where we show the cosmic evolution of optically selected AGN in 11 magnitude bins (left panel) and soft X-ray selected AGN in 8 luminosity bins (right panel). We have depicted separately the evolution of the total number of AGN (before applying obscuration; dashed lines) and visible population of AGN (after applying obscuration; solid lines) in each bin. In this figure we also show the sum over all magnitudes with $\Mbj\leqslant-18$ and luminosities with $\log L_{\rm{SX}}\geqslant10^{42}\ergsec$ (black solid lines for the visible and black dashed lines for the total populations).

A feature immediately evident in both wavebands is that the evolution of the space density of AGN is shallower for fainter sources. For example, in the optical waveband the increase in the space density of the $\Mbj=-21$ AGN is characterized in the redshift interval $z=0-1$ by a power law of the form $(1+z)^\beta$ with a very shallow slope of $\beta\sim0.5$. The slope becomes gradually steeper reaching a value of $\sim2.3$ for the $\Mbj=-26$ AGN. The steepening of the slope is evident in both the total and visible populations and it is driven primarily by the two distinct accretion channels in our model as follows. The faintest sources depicted in the plots of Fig.~\ref{downsizing} comprise a combination of objects accreting during the starburst and hot-halo mode. The hot-halo mode is responsible for building up the majority of them in the low redshift universe whereas the starburst mode accounts for them exclusively in the high redshift universe (cf. Figs.~\ref{ledd} and \ref{mbh_lbol}, but also Fig.~\ref{blf}). The slope of these populations changes very slowly since the transition from the hot-halo mode to starburst-mode domination is not characterized by a strong change in the space density of AGN (the decrease of hot-halo sources is compensated by the increase of starburst sources).  However, when we consider brighter populations, the contribution of the hot-halo mode becomes less important (or vanishes for the brightest populations). As a consequence, in the low redshift universe we find a much steeper slope for the bright AGN since their evolution is driven primarily by the starburst mode. 

A second noticeable feature is the strong reduction of the visible populations compared to the total AGN populations. The effect is stronger in the faintest populations where we see a decrease in the space density of AGN sometimes greater than an order of magnitude. This has a significant effect on the redshift at which the space density of AGN peaks. For example, the space density of the $\Mbj\leqslant-18$ and $\log L_{\rm{SX}}\geqslant10^{42}\ergsec$ AGN (dashed black lines) peaks at $z\simeq3.5$, which is the same redshift where the SF in bursts peaks (Fig.\ref{sfr}). The same behaviour is seen also in the individual luminosity AGN populations in both the optical and soft X-rays. However, the visible population shows a different behaviour. AGN activity, as observed in the soft X-rays for example, seems to peak at $z\sim2-2.5$, much later than that of the total AGN population. More interestingly, when considering each luminosity population separately we notice that the brightest sources peak at $z\sim3$ whereas the faintest ones at $z\lesssim2$. Similarly, although less obviously, in the optical the space density of the faintest quasars peaks at $z\sim2$ whereas that of the most luminous increases monotonically until $z\simeq3$. This behaviour is ascribed to the fact that fainter populations are subject to stronger obscuration. Therefore their abundance reaches a maximum at lower redshifts and beyond that it declines faster. The strong obscuration also contributes to the further flattening of the slope making the evolution of these AGN quite modest.

Trends similar to those seen in the cosmic evolution of AGN in our model have been reported in the literature. For example, \citet{hopkins_2007} find in their analysis of the bolometric LF that faint AGN evolve much more slowly than bright AGN (Fig.~9 in their analysis). The modest evolution is evident also in the optical and soft/hard X-rays. In addition, \citet{hasinger_2005} report that in their sample of soft X-ray AGN, the redshift at which the space density of AGN peaks changes significantly with luminosity: the peak of AGN with $10^{42}\ergsec<L_{\rm{SX}}<10^{43}\ergsec$ is at $z\simeq0.7$ but gradually shifts to higher redshifts, reaching $z\simeq2$ for $10^{45}\ergsec<L_{\rm{SX}}<10^{46}\ergsec$. Similarly, \citet{croom_2009b} found the same downsizing behaviour in the optical sample of quasars from 2SLAQ+SDSS. The faintest quasars in their sample peak at $z\sim0.6-0.8$ ($-21.5\geqslant M_{g}\geqslant-22.5$) whereas the brightest sources ($M_{g}\leqslant-25.5$) seem to monotonically increase in density up to $z\simeq2.5$. Further evidence that the redshift at which the space density of quasar peaks is a strong function of luminosity form the analysis of \citeauthor{croom_2009b} is the inability of pure luminosity evolution  models (PLE; models where only the luminosity of objects changes) to provide an excellent fit to the data. Instead, luminosity and density evolution (LEDE; models where both the luminosity and density of objects change) are needed to provide the best fit to the data. 

These predictions pose the following key question. Why does AGN activity shift from high-luminosity objects in the high-$z$ universe to low-luminosity objects in the low-$z$ universe? Is this clear evidence against hierarchical galaxy formation models as has been interpreted by many authors? In our model, the downsizing of AGN is expressed through the shallow slope that characterizes the evolution of faint AGN and the dependence on luminosity of the redshift at which the AGN density in a given luminosity range peaks. The former arises naturally from the interplay of the starburst and hot-halo mode, just as in the galaxy population \citep{bower_2006}. Therefore, the SF and cooling processes in a hierarchical cosmology can account consistently for the different evolution with redshift of the individual luminosity populations. The latter is driven largely by obscuration biases rather than the nature of their host galaxies. The analysis presented earlier in this section shows clearly that the cosmic evolution of the AGN space density is subject to strong obscuration which depends significantly on both luminosity and redshift. Therefore, AGN populations that are subject to obscuration could lead to misleading conclusions. For example, in both the \citet{croom_2009b} QSO (these are only type-1 AGN \citet{croom_2009a}) and \citet{hasinger_2005} soft X-ray AGN samples no correction for obscuration is adopted. Given that at low redshifts only unabsorbed AGN are detected in soft X-rays \citep{lafranca_2005}, these samples can only trace the evolution of a small fraction (and also a specific class) of the total AGN population. 

Hence, to probe the intrinsic evolution of AGN one needs to correct for observational biases caused by obscuration (note also that incompleteness effects at low luminosities and high redshifts could also influence observational results about the evolution of AGN). \citet{ueda_2003} investigated the cosmological evolution of hard X-ray AGN  using a combination of surveys such as the HEAO 1, ASCA and Chandra. The sample of AGN constructed by \citeauthor{ueda_2003}, which was compared to our predictions for the hard X-ray LF in Section~\ref{sec:The X-ray LFs}, was corrected using an absorption distribution function that depends on both the luminosity and redshift. In a similar analysis, \citet{lafranca_2005} used a sample of AGN from the HELLAS2XMM survey to study the evolution of the hard X-ray LF taking into account selection effects due to X-ray absorption. In both these studies, a careful examination of the ``cosmic evolution" plots (Fig.~12 in \citeauthor{ueda_2003} and Fig.~8 in \citeauthor{lafranca_2005}) reveals no evidence for downsizing in hard X-rays. 

When accounting for these effects, our model suggests that the cosmic interplay between the starburst and hot-halo mode results in complex evolution scheme for the AGN. In this scheme, low luminosity AGN avoid the cosmic fate (namely becoming dramatically less numerous with time) of their bright counterparts, since accretion during the hot-halo mode provides gas for enabling modest AGN activity in the low-$z$ universe. Note that, as already implied by our analysis, modest activity implies low accretion rates rather than low BH masses. The AGN in the hot-halo mode show a wide range of BH masses, and therefore, in our model the AGN downsizing \emph{does not} imply that the growth of low-mass BHs is delayed to low redshifts. Averaged-sized BHs accreting at low rates as an interpretation of the reported AGN downsizing is also supported by observations of X-ray selected AGN in the Chandra Deep Field South at $z < 1$ \citep{babic_2007}.

\section{Conclusions}

In this paper we have made predictions for the evolution of AGN using an extended version of the semi-analytic code \texttt{GALFORM}. \texttt{GALFORM} simulates the formation and evolution of galaxies in a $\Lambda$CDM cosmology. In \citet{fanidakis_2010}, we introduced a calculation of BH spin. The fiducial model in this paper uses an improved SF law, as implemented by Lagos \etal (2010). 

Using this model we calculate the cosmic evolution of the fundamental parameters that describe BHs in our model. These are the BH mass, $\Mbh$, BH spin, $a$, and accretion rate onto the BH, $\dot{m}$ (expressed in units of the the Eddington accretion rate through out this paper). We find that at high redshifts ($z\sim6$) it is mainly the $10^{6}-10^{7}\Msun$ BHs accreting at $\dot{m}\simeq0.3$ that are actively growing. This picture changes at lower redshifts where we find that the accretion activity peaks for $10^{7}-10^{8}\Msun$ BHs, accreting at $\dot{m}\simeq0.05$. Throughout the evolution of these BHs their spin is kept low because of the chaotic fashion with which gas is consumed. However, when these BHs grow to masses of $\gtrsim5\times10^{8}\Msun$ they acquire high spins as a consequence of mergers with other BHs. 

Knowledge of the values of $\Mbh$, $a$ and $\dot{m}$ allows us to calculate the luminosity produced during the accretion of gas. To do so, we assume that accretion takes place in two distinct regimes: the thin-disc (radiatively efficient) and ADAF (radiatively inefficient) regime. Objects accreting in the thin-disc regime are luminous enough to account for the observed luminosities of AGN. Using the luminosities predicted for each AGN (formed when the central BH experiences an accretion episode) and a prescription for taking into account the effects of obscuration \citep{hasinger_2008} we calculate the LF of all accreting objects in the optical, soft and hard X-rays. We find an a very good agreement with the observations, particularly in the optical, by adjusting the value of the $\fq$ parameter, namely the proportionality factor that determines the accretion timescale in our model. 

The presence of two distinct accretion channels naturally causes a downsizing in the predicted AGN populations. These channels shape the evolution of the faint AGN with cosmic time, indicating that the faint end of the LF is dominated by massive BHs experiencing quiescent accretion. By contrast, the bright end is always populated by AGN radiating near or greater than the Eddington limit. In addition, our model suggests that a significant fraction of AGN are obscured in optical and soft X-rays. In fact, we find that $\sim90\%$ of the total number of AGN in the $z=0.015-4.8$ universe are not visible in soft X rays. The implications of the obscuration are further revealed when we study the cosmological evolution of AGN of different luminosity populations. As demonstrated by Fig.~\ref{downsizing}, low luminosity AGN populations are strongly attenuated by obscuration. As a result, their peak of AGN activity appears shifted to lower redshifts. 

This analysis explicitly demonstrates that hierarchical cosmological models for galaxy formation and evolution are able to provide a robust framework in which the evolution of AGN can be studied. The ability of our galaxy formation model to reproduce the observed LFs and account for the mechanisms that shape the evolution of the faint and bright AGN populations strengthens the powerful capabilities of semi-analytic modelling and shows that the level of AGN activity implied by AGN feedback is compatible with observations. In future studies, we aim to further study the statistical properties of AGN by exploring their spatial clustering and environmental dependence.

\section*{Acknowledgements}
NF acknowledges receipt of a fellowship funded by the European Commission's Framework Programme 6, through the Marie Curie Early Stage Training project MEST-CT-2005-021074. AJB acknowledges the support of the Gordon \& Betty Moore Foundation. SC acknowledges the support of a Leverhulme research fellowship. CSF acknowledges a Royal Society Wolfson Research Merit Award. This work was supported in part by an STFC Rolling Grant to the Institute for Computational Cosmology.

\bibliographystyle{mn2e}

\end{document}